\documentstyle[12pt,worldsci,epsf,subfigure]{article}
\newcommand{\postscript}[2]
 {\setlength{\epsfxsize}{#2}
  \centerline{\epsfbox{#1}}}
\newlength{\mswsize}
\begin{document}

\title{
                     SOLAR NEUTRINOS: HINT FOR NEUTRINO MASS
		     \footnote{Based on a talk given at Solar Modeling
			       Workshop, Institute of Nuclear Theory,
			       University of Washington, Seattle,
			       March 24, 1994.}
}
\author{
                     Naoya Hata	                               \\
{\em
                     Department of Physics,
                     University of Pennsylvania,               \\
                     Philadelphia, Pennsylvania 19104, USA    \\ 
}
                     (April 22, 1994, UPR-0612T)                 
}
\maketitle
%

\begin{abstract}

The present status of the solar neutrino problem is reviewed.  The
strongest motivation for new neutrino physics comes from the complete
phenomenological failure of astrophysical solutions: (1) The standard
solar model is excluded by each of the solar neutrino experiments. (2)
The combined results of Homestake and Kamiokande are incompatible with
any astrophysical solution. (3) Even if the Homestake results were
ignored entirely, there is no realistic solar model so far that
describes the Kamiokande and gallium data simultaneously within their
experimental uncertainties.  On the other hand, the MSW solutions
provide a complete description of the data, strongly hinting at the
existence of mass and mixing of neutrinos.  The MSW predictions for
the future experiments are given in detail.  It is stressed that solar
model uncertainties are significant in determining the oscillation
parameters and in making predictions for the future experiments.
Especially, a $^8$B flux (or $S_{17}$) smaller by 20\% than that of
the Bahcall-Pinsonneault model can make the interpretation of the
charged to neutral current ratio measurement in SNO ambiguous.

\end{abstract}


\section{		Introduction	}
\label{sec_intro}

There is a mounting evidence that the solar neutrino flux is
considerably less than the theoretical predictions of the standard
solar model (SSM).  The SSM predictions by Bahcall and Pinsonneault
\cite{Bahcall-Pinsonneault} and by Turck-Chi\`eze and Lopes
\cite{Turck-Chieze-Lopes} are listed in Table~\ref{tab_expdata} along
with the experimental results.  The recent SSM by Bahcall and
Pinsonneault uses the updated input parameters and includes the helium
diffusion effect.  The observed rates of the Homestake chlorine
experiments \cite{Homestake,Homestake-update}, Kamiokande water
\v{C}erenkov experiment \cite{Kamiokande-II,Kamiokande-III}, and the
combined result of the SAGE \cite{SAGE,SAGE-update} and GALLEX
\cite{GALLEX} gallium experiments are 3.1, 5.5, and 4.6 standard
deviations smaller than their predictions, including the theoretical
uncertainties.  The deficit is 1.3, 2.9, and 3.7 $\sigma$,
respectively, when the data are compared to the Turck-Chi\`eze--Lopes
SSM.  The differences of the two calculations are understood as
differences of input parameters, and also the Turck-Chi\`eze--Lopes
model does not include the helium diffusion effect and assigns more
conservative uncertainties.

\begin{table}[hbt]
\caption{
%
%
The standard solar model predictions of Bahcall and Pinsonneault (BP SSM)
\protect\cite{Bahcall-Pinsonneault} and of Turck-Chi\'eze and Lopes
(TL SSM) \protect\cite{Turck-Chieze-Lopes}, along with the results of
the solar neutrino experiments.
}
\label{tab_expdata}
\vspace{1.0ex}
\scriptsize
\begin{tabular}{l  c c c}
\hline
\hline
\\
               & BP SSM        & TL SSM       & Experiments \\[2ex]
\hline
\\
Kamiokande$^a$ &  1 $\pm$ 0.14 & 0.77 $\pm$ 0.19 & $(0.51\pm0.07) \times
						 5.69$E6 /cm$^2$/sec (BP SSM)
							\\[2ex]
Homestake$^b$ (Cl)&  8 $\pm$ 1 SNU  & 6.4 $\pm$ 1.4 SNU & 2.32 $\pm$ 0.26 SNU
                                                 (0.29 $\pm$ 0.03 BP SSM)
							\\[2ex]
SAGE$^c$ \& GALLEX$^d$ (Ga)
		& 131.5 $^{+7}_{-6}$ SNU & 122.5 $\pm$ 7 SNU & 78 $\pm$ 10 SNU
                                                    (0.59 $\pm$ 0.08 BP SSM)
							\\[2ex]
\hline
\hline
\normalsize
\end{tabular}
{\footnotesize
$^a$ The combined result of Kamiokande II
[ 0.47 $\pm$ 0.05 (stat) $\pm$ 0.06 (sys) BP SSM, 1040 days]
\cite{Kamiokande-II} and Kamiokande III
[ 0.55 $\pm$ 0.06 (stat) $\pm$ 0.06 (sys) BP SSM, 627 days]
\cite{Kamiokande-III} is
0.51 $\pm$ 0.04 (stat) $\pm$ 0.06 (sys) BP SSM
\cite{Kamiokande-III}.  The $^8$B flux in the Bahcall-Pinsonneault (BP)
SSM with the helium diffusion effect is 5.69$\times$10$^6$ /cm$^2$/sec. \\
$^b$ The result through June, 1992 (Run 18 -- 124) is
2.32 $\pm$ 0.16 (stat) $\pm$ 0.21 (sys) SNU \cite{Homestake-update}. \\
$^c$ The preliminary result of SAGE I (through May, 1992) is
74 $\pm$ 19 (stat) $\pm$ 10 (sys) SNU \cite{SAGE-update}. \\
$^d$ The combined result of GALLEX I and II (30 runs, through October, 1993)
is 79 $\pm$ 10 (stat) $\pm$ 6 (sys) SNU \cite{GALLEX}.
}
\end{table}

\begin{figure}[hbt]
\begin{center}
\begin{tabular}{c}
\setlength{\epsfxsize}{0.45\hsize}
\subfigure[Bahcall-Pinsonneault SSM]{\epsfbox{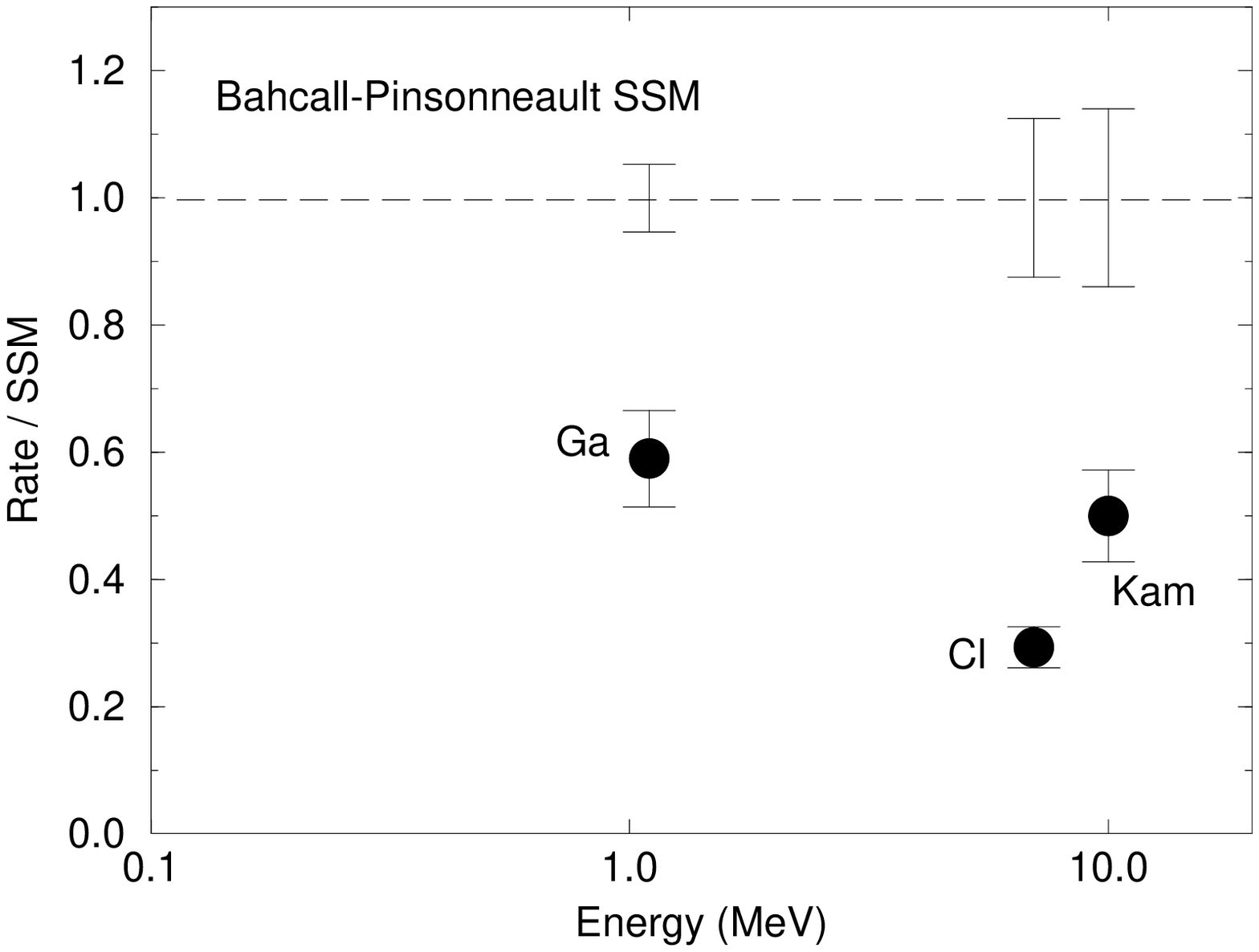}}
\hspace{5mm}
\subfigure[Turck-Chi\`eze--Lopes SSM]{\epsfbox{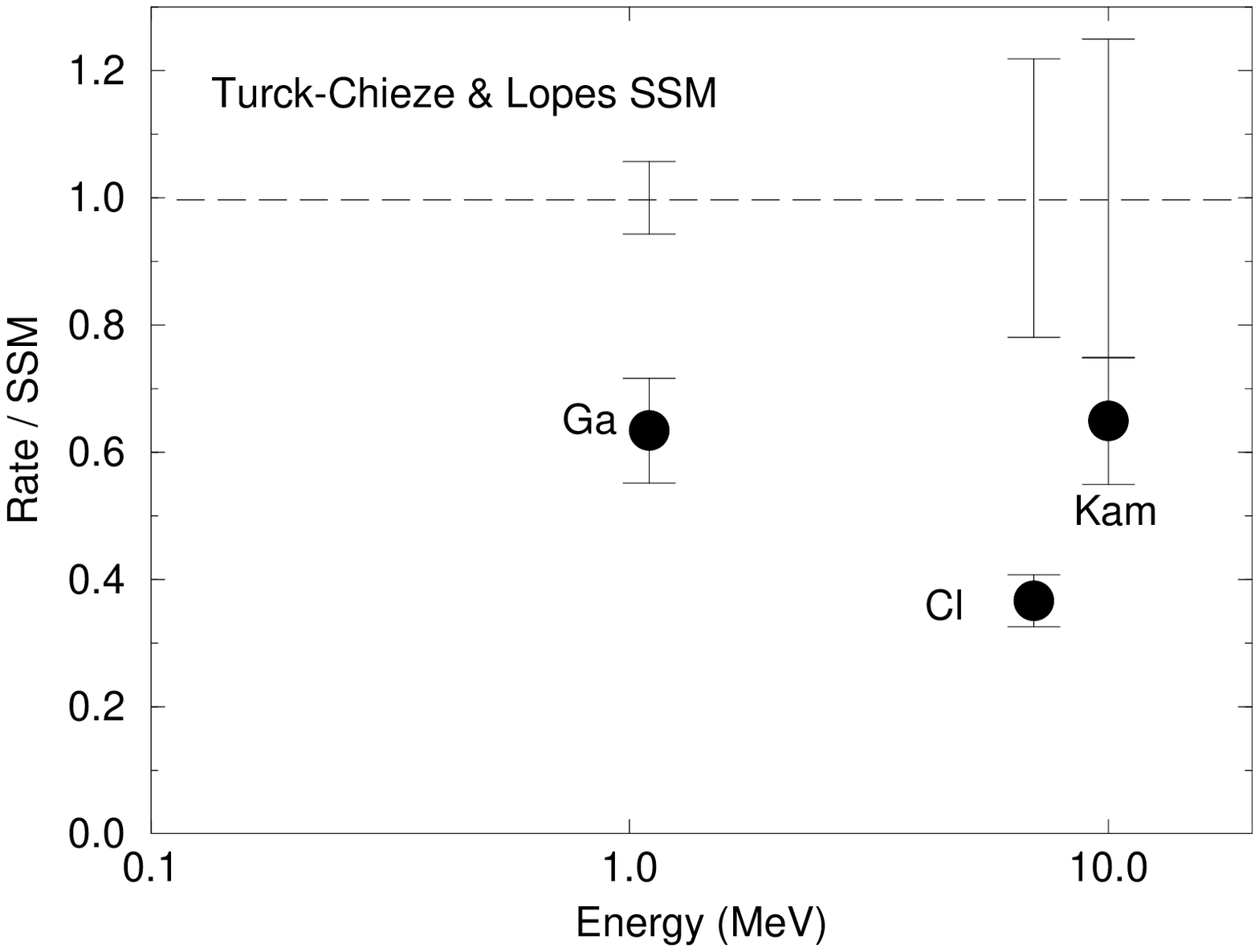}}
\end{tabular}
\end{center}
\caption{
%
%
A graphical summary of the solar neutrino problem.  The experimental
rates of the combined SAGE and GALLEX, Homestake, and Kamiokande
experiments are shown with respect to (a) the Bahcall-Pinsonneault SSM
and (b) the Turck-Chi\`eze--Lopes SSM.  The data points are plotted
according to the energy sensitivities of the experiments, but only
crudely.  The theoretical uncertainties are also displayed as error
bars (1$\sigma$).
}
\label{fig_expdata}
\end{figure}

The solar neutrino data are graphically compared to the SSM
predictions in Fig.~\ref{fig_expdata}.  The solar neutrino problem is
summarized in two aspects:
\begin{itemize}

\item	Assuming the standard properties of neutrinos, the SSM is
	excluded.  Neither the Bahcall-Pinsonneault or the
	Turck-Chi\`eze--Lopes model (or any other SSM) are consistent
	with the observations.

\item   More importantly, the smaller observed rate in the Homestake
	data relative to Kamiokande is inconsistent with astrophysical
	solutions in general.  When the neutrino flux is reduced by
	changing the solar model, the $^8$B flux is generally reduced
	more than the $^7$Be flux, and therefore the Kamiokande rate
	should be suppressed more than the Homestake rate, contrary to
	the data.

\end{itemize}

{}From a particle physicist's point of view the most exciting
interpretation of the problem is that the solar neutrino deficit is a
manifestation of nonstandard neutrino properties, such as mass and
mixing assumed in the Mikheyev-Smirnov-Wolfenstein (MSW) mechanism.
This possibility is, however, one of three generic possibilities, and
any combination of the three is possible.
\begin{itemize}

\item	{\it Experimental Solutions:}  One or more of the experiments is
	wrong.

\item 	{\it Astrophysical Solutions:}  The SSM is wrong.  Nonstandard
	input parameters and/or nonstandard processes that are not
	included in the SSM will explain the deficit of the solar
	neutrino flux.

\item	{\it Particle Physics Solutions:}  The standard model of the
	electroweak theory is wrong.  Nonstandard neutrino properties
	are causing the deficit.  Proposed solutions include the MSW
	effect, vacuum oscillations, a large neutrino magnetic moment,
	neutrino decays, violation of the equivalence principle, etc.

\end{itemize}
In this paper I will focus my discussion on the theoretical aspects of
the solar neutrino problem.  For astrophysicists what is at stake is
the understanding of the Sun, which is the best measured main sequence
star.  For particle physicists, it is new neutrino physics such as
neutrino mass and mixing.  These neutrino properties require a
(long-waited) extension of the standard model and provide more
information on the fermion mass spectrum and fermion mixing, whose
theoretical understanding is totally lacking.

It is not my intention here to discuss the reliability of the
experiments explicitly other than mentioning that no defect has been
found so far in any of the existing experiments.  Instead I will use
the experimental data as given and will quantitatively test various
theoretical ideas.  I also address my objection to the theoretical
analyses in the literature done by selecting a subset of available
data set (e.g., discarding the earlier Homestake data).  This kind of
data selection is unfounded and misleading, and often defocuses the
issue we are facing.

This paper is organized as the following.  In Section~\ref{sec_astro},
astrophysical solutions are discussed for the cool Sun model and for
the model independent analysis.  In fact the most compelling reason
for considering particle physics solutions comes from the complete
failure of astrophysical solutions to describe the data.  It will be
shown that the observed lower Homestake rate relative to the
Kamiokande rate essentially excludes all astrophysical solutions.  The
lack of a viable astrophysical explanation motivates us to consider
particle physics solutions.  In Section~\ref{sec_msw}, the MSW
solutions are discussed as the simplest particle physics solution.
The results of the updated MSW analysis are presented.  The analysis
includes the Earth effect, the Kamiokande II day-night data, the SSM
theoretical uncertainties and their correlations, a proper joint
analysis of the experiments, and an improved determination of the
confidence levels.  Assuming the SSM, the experiments allow two
parameter regions at 95\% C.L.: one in the small-angle (nonadiabatic)
and one in the large-angle region, the former solution giving a better
fit.  There is also a small-angle (nonadiabatic) solution for
oscillations to sterile neutrinos, but none in the large-angle region
even at 99\% C.L.  The MSW effect can also be considered in
nonstandard solar models.  The allowed parameter space is obtained for
nonstandard core temperatures, nonstandard $^8$B fluxes, and various
explicitly constructed nonstandard solar models.  It is stressed that,
once the MSW effect is established in the next generation solar neutrino
experiments, the uncertainties in the neutrino parameters will be
dominated by the solar model uncertainties, and the calibration of the
neutrino flux in the presence of the MSW effect will be a major
theoretical issue.  Assuming the SSM and the existing data are
correct, the MSW effect provides robust predictions for the
next-generation experiments such as SNO, Super-Kamiokande, BOREXINO,
and ICARUS, and those predictions are given in detail in
Section~\ref{sec_msw}.  I argue that the prediction for the charged to
neutral current ratio in SNO is solar model dependent, and a correct
estimation of the $^8$B flux uncertainty is crucial.


\section{		Astrophysical Solutions		}
\label{sec_astro}

The SSM calculations have been refined over the years, and the recent
model by Bahcall and Pinsonneault improved on earlier calculations by
incorporating the new Livermore opacity and updated element
abundances, and also by including the helium diffusion effect for the
first time in the SSM calculation \cite{Bahcall-Pinsonneault}.  The
model is in good agreement with the helioseismology observations.
They also showed that the various SSM calculations give essentially
the same flux predictions if the same input parameters were used,
proving the reliability of the calculations within the SSM framework.
More recent attempts to incorporate mixing processes, heavy element
diffusion, and rotations were reported by other speakers at this
workshop \cite{Charbonnel,Pinsonneault,Demarque,Glasner,Stix}.

Despite those successes and improvements of the SSM, one has to depart
from the SSM in order to search for an astrophysical solution of the
solar neutrino problem.  It is well known that the neutrino flux
prediction responds sensitively to the changes in some of the solar
model input parameters.  For example, opacity or heavy element
abundance smaller than the standard values can achieve lower neutrino
fluxes.  Astrophysical processes that are not included in the SSM
calculations, such as rotations, core magnetic fields, or
gravitational settling of elements, might significantly reduce the
neutrino flux.  Those effects usually work by reducing the core
temperature, and the neutrino fluxes are sensitive to the change in
the core temperature.

Low energy nuclear reaction cross sections are another source of flux
uncertainties.  The low energy cross sections important in the SSM
such as $S_{17}$, $S_{33}$, $S_{34}$, and $S_{11}$ have never been
directly measured at the low energy equivalent to the solar
temperature, and there could still be unquantified theoretical
uncertainties in extrapolating the measured cross sections to the
lower energies.  Such uncertainties might be correlated among
different cross sections.

In particular $S_{17}$, to which the $^8$B flux magnitude is directly
proportional, is a poorly measured quantity, and I believe that its
quoted uncertainty is underestimated.  The standard value ($S_{17}(0)
= 22.4 \pm 2.1 \; \mbox{eV b}$) recommended in
Ref.~[\citenum{Johnson-etal,Langanke}] is the combined result of six
measurements, but is controlled mostly by the two inconsistent
measurements with relatively small uncertainties: $S_{17}(0) = 25.2
\pm 2.4 \; \mbox{eV b}$ of Ref.~[\citenum{Kavanagh-etal}] and $20.2
\pm 2.3 \; \mbox{eV b}$ of Ref.~[\citenum{Filippone-etal}].  From
the differential cross section plot of the two data shown in
Ref.~[\citenum{Johnson-etal}], it is clear that the two measurements
are inconsistent: if one result is correct, the other must be wrong.
The true value can be $20.2 - 2.3 = 17.9 \; \mbox{eV b}$ or $25.2 +
2.4 = 27.6 \; \mbox{eV b} $.  My objection for the combined value is
that its uncertainty does not cover the entire possible range.  As I
will show in Section~\ref{sec_model-ind}, the lower choice of $S_{17}$
does not solve the solar neutrino problem.  However, underestimating
the uncertainty does change the obtained MSW parameters, and therefore
change the MSW predictions for the future experiments.  The charged to
neutral current ratio experiment in SNO is the gold-plated measurement
to establish neutrino oscillations, but, given the currently projected
systematic uncertainty of 20\%, a $^8$B (or $S_{17}$) lower by 20\%
than the Bahcall-Pinsonneault value can make the conclusion of the
measurement ambiguous.

The worst scenario is that those astrophysical effects and nuclear
physics effects might be taking place at the same time.  In fact
changes in nuclear reaction cross sections such as $S_{11}$ can cause
a change in the core temperature.  This complexity and uncertainty of
the problem allow theorists to come up with new astrophysical solutions
now and then, such as a recent claim of the standard solar model
solution to the solar neutrino problem
\cite{Dar-Shaviv}.  (Fatal flaws of the claim are discussed elsewhere
\cite{Bahcall-etal,Bahcall,Parker,Langanke}.)

Despite all those worrisome aspects of the solar model calculation, it
can be shown that the data cannot be explained by astrophysical and
nuclear effects.  In the following, I will discuss two analyses of
nonstandard solar models: the cool sun model, following the analysis
of Ref.~[\citenum{BKL,BHKL}] (see also
Ref.~[\citenum{Bahcall-Bethe,GALLEX,Shi-Schramm-PW,Degl'Innocenti}]),
and the model independent analysis, following Ref.~[\citenum{HBL}],
which depends essentially only on the data and provides powerful
conclusions.  The cooler Sun analysis shows that the data exclude a
wide class of nonstandard solar models characterized by lower core
temperatures; the model independent analysis shows that, under minimal
assumptions, the data are incompatible with any astrophysical/nuclear
solutions.


\subsection{		Cool Sun Model	  }
\label{sec_cool-sun-model}

A class of nonstandard solar models can be characterized by lower core
temperatures \cite{BKL,BHKL}.  The validity of this general
description was explicitly demonstrated by Degl'Innocenti for solar
models with large $S_{11}$, low opacity, low Z, and low age
\cite{Degl'Innocenti}.  The neutrino fluxes are described by the power
laws of the central temperature ($T_C$) obtained from 1000 Monte-Carlo
SSMs\cite{Bahcall-Ulrich,Bahcall-book}:
\begin{equation}
\label{eqn_Tc-power-law}
        \phi(pp)        \sim T_C^{-1.2}, \;
        \phi(\mbox{Be}) \sim T_C^{8},   \; \mbox{ and } \;
        \phi(\mbox{B})  \sim T_C^{18},
\end{equation}
Note that these relations are obtained within the SSM calculation,
and there are uncertainties in the exponents in the power law,
especially when $T_C$ is changed beyond the SSM uncertainty ($T_C = 1
\pm 0.006$).  (The exponents obtained by Degl'Innocenti is $-$0.6, 9,
and 21 for the $pp$, $^7$Be, and $^8$B flux, respectively
\cite{Degl'Innocenti}.)  Especially, the power law for the pp flux is
only valid within the SSM uncertainty, and a more reliable $T_C$
dependence is obtained by the luminosity constraint, which determines
the total number of neutrinos generated by nuclear reactions.  Taking
account the energies carried off by neutrinos, the luminosity
constraint yields
\begin{equation}
\label{eqn_luminosity}
	  \phi(pp) + 0.979 \, \phi(\mbox{Be}) + 0.955 \,  \phi(\mbox{CNO})
	= 6.57 \times 10^{10} \, \mbox{cm}^{-2}\mbox{s}^{-1},
\end{equation}
where the fluxes are described as functions of $T_C$.  We use the
power law (Eqn.~\ref{eqn_Tc-power-law}) for the $^7$Be and $^8$B flux
and take the exponent of the CNO flux as 22 \cite{HL-MSW-analysis}.
$\phi(pp)$ is then obtained from Eqn.~\ref{eqn_luminosity}.  The
central temperature is in units of the Bahcall Pinsonneault SSM value
($T_C = 1 = 15.57\times10^{6} \; \mbox{K}$).  Using the $T_C$
dependence, one can write the expected rates of the Kamiokande
($R_{\mbox{\scriptsize Kam}}$), chlorine ($R_{\mbox{\scriptsize
Cl}}$), and gallium ($R_{\mbox{\scriptsize Ga}}$) experiments relative
to the SSM predictions as functions of $T_C$:
\begin{eqnarray}
    R_{\mbox{\scriptsize Kam}}
                &=&  (1 \pm 0.089) \, T_C^{18}                  \\ \nonumber
    R_{\mbox{\scriptsize Cl}}
                &=& (1 \pm 0.033) \;
                  [ \, 0.775 \, (1 \pm 0.089) \, T_C^{18}
                   +0.150 \, (1 \pm 0.036) \, T_C^8 + \hbox{small terms} \; ]\\
    R_{\mbox{\scriptsize Ga}}
                &=&  (1 \pm 0.04) \; [ \,
                   0.538{\phi(pp) \over \phi(pp)_{\mbox{\scriptsize SSM}} }
		+ 0.272 \, (1 \pm 0.036) \, T_C ^8
  							\\ \nonumber
         & & 
		+ 0.105 \, (1 \pm 0.089)\,  T_C^{18}
		+ \hbox{small terms} \;  ].                  \nonumber
\end{eqnarray}
Our analysis includes the uncertainties (shown in parentheses) and
their correlations from the nuclear reaction cross sections and the
detector cross sections.

When the data are fit to $T_C$ separately, the Kamiokande and
Homestake data require $T_C = 0.96 \pm 0.01$ and $0.92 \pm 0.01$,
respectively.  The gallium rate of $78 \pm 10$ SNU requires $T_C =
0.71 \pm 0.14$, whose central value is unrealistically low compared to
the SSM value ($T_C = 1 \pm 0.006$).  Moreover, the temperatures
obtained are inconsistent between different experiments.  When all
data are fit simultaneously, the best value is $T_C = 0.93 \pm 0.01$,
but only with an extremely poor fit; $\chi^2 = 15.7$ for 2 degrees of
freedom, and statistically this model is excluded at 99.96\% C.L.  The
results of $T_C$ fit to various combinations of the data are listed in
Table~\ref{tab_Tcfit}, along with the expected rate for each
experiment from the obtained temperatures.  It should also be noted
that there is no consistent temperature even if we ignore one of the
three experiments.

\begin{table}[hbt]
%
%
\caption{
The results of the $T_C$ fits.  The expected rates for the three
experiments are listed for each obtained $T_C$; the Kamiokande rate is
in units of the Bahcall-Pinsonneault SSM and the chlorine and gallium
rates are in units of SNU.}
\label{tab_Tcfit}
\vspace{1.0ex}
\begin{center}
\begin{tabular}{l  r r r r r r}
\hline \hline 
		&$T_C\pm\Delta T_C$ &$\chi^2$&C.L. (\%)&Kam & Cl & Ga \\
\hline 
Kam		& 0.96 $\pm$ 0.01 & 0    & --   & 0.51 & 4.2 & 111  \\
Cl		& 0.92 $\pm$ 0.01 & 0    & --   & 0.22 & 2.3 &  99  \\
Ga              & 0.71 $\pm$ 0.14 & 0.08 & --   & 0.002& 0.16&  81  \\
Kam+Cl		& 0.93 $\pm$ 0.01 & 10.6 & 99.98& 0.29 & 2.7 & 102  \\
Kam+Ga		& 0.96 $\pm$ 0.01 & 9.0  & 99.7 & 0.44 & 3.8 & 109  \\
Cl+Ga           & 0.92 $\pm$ 0.01 & 4.2  & 96   & 0.23 & 2.2 &  99  \\
Kam+Cl+Ga       & 0.93 $\pm$ 0.01 & 15.7 & 99.96& 0.28 & 2.6 & 102  \\
\hline 
\hline 
\end{tabular}
\end{center}
\end{table}

We have reanalyzed the data with various other combinations of the
exponents of the $T_C$ power laws, and have found that the as long as
the $^8$B flux is more temperature sensitive than the $^7$Be flux, the
combined observations cannot be described by  lower core temperatures.
The fit has been repeated with all theoretical uncertainties tripled.
The result is somewhat modified, but the conclusion remains the same:
the cool suns do not explain the solar neutrino observations.


\subsection{		Model Independent Analysis	  }
\label{sec_model-ind}

A more powerful conclusion can be obtained just from the observations,
without referring to any specific nonstandard solar models.  Here I
follow the analysis of Ref.~[\citenum{HBL}], but a similar approach
can be found in
Ref.~[\citenum{Castellani-Degl'Innocenti-Fiorentini-AA}].

Suppose we do not know anything about the theory of the Sun; what will
the solar neutrino data tell us about the star?  We have formulated
the question by using four relevant neutrino fluxes ($pp$, $^7$Be,
$^8$B, and CNO) as free parameters fit to the observations.  Except
for the luminosity constraint, there are no relations imposed among
the fluxes as in the cool sun model.  It is not my intention to
advocate such a model as a realistic description of the Sun or to
claim that it is consistent with other observables, but to illuminate
the problems of the astrophysical solutions when they are confronted
by the observations.

While using the four fluxes as free parameters, we made three explicit
assumptions:
\begin{enumerate}
\item	The Sun is in a quasi-static state, and the solar luminosity
	is generated by the ordinary nuclear fusions.  That is, the
	Sun shines as it constantly consumes its nuclear fuel through
	the $pp$ and CNO chains.  This imposes a relation among the
	fluxes (Eqn.~\ref{eqn_luminosity}).

\item	Astrophysical mechanisms do not distort the neutrino energy
	spectrum at the observable level.  We are allowed to change,
	for example, the amplitude of the $^8$B flux freely, but not
	to deplete only the lower energy part of the spectrum in order
	to reconcile the Homestake and Kamiokande rates.  It was shown
	by Bahcall that possible distortions of the spectrum due to
	such astrophysical effects as gravitational red-shifts and
	thermal fluctuations are completely negligible
	\cite{Bahcall-spectrum}.  Note that particle physics solutions
	are often energy dependent and do distort the spectrum: the
	MSW small-angle (nonadiabatic) solution is a typical example.

\item	The experiments are correct, and so are the detector cross section
	calculations.
\end{enumerate}

\begin{figure}[htb]
\postscript{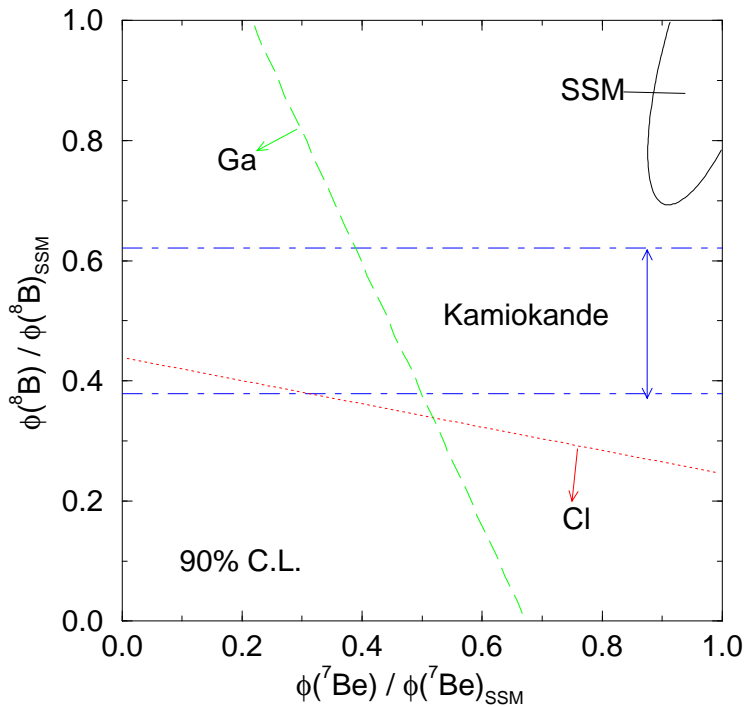}{0.70\hsize}                            
\caption{
%
%
The allowed fluxes when the Homestake, Kamiokande, and the gallium
results are fit separately.  This is an updated analysis of
Ref.~[\protect\citenum{HBL}].
}
\label{fig_fff-each}
\end{figure}

\begin{figure}[hbt]
\postscript{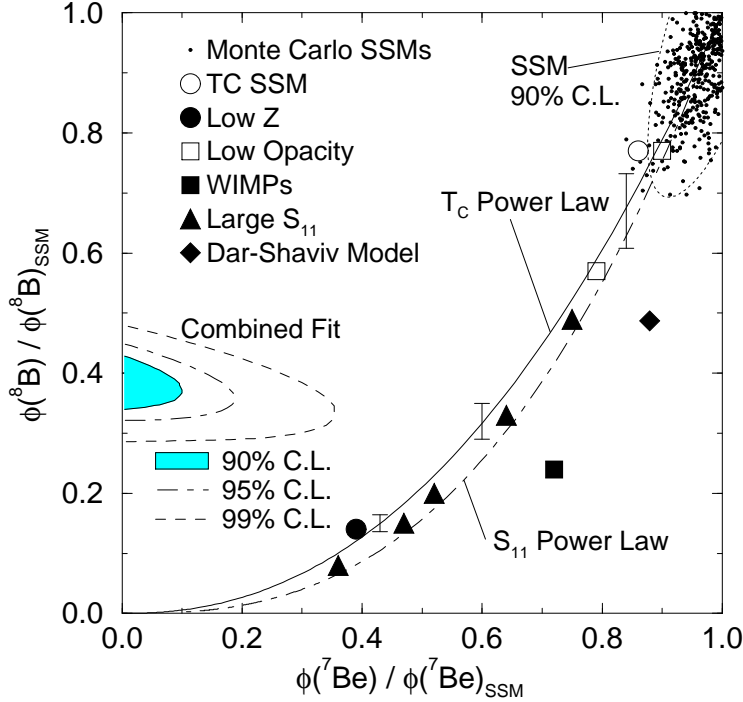}{0.70\hsize}                            
\caption{
%
%
The fluxes allowed by the combined data.  Also shown are various
standard and nonstandard solar models.  This is an updated analysis of
Ref.~[\protect\citenum{HBL}].
}
\label{fig_fff-comb}
\end{figure}

The results of the analysis are presented in the 2-dimensional plane
of the $^7$Be and $^8$B flux in units of the Bahcall-Pinsonneault SSM
values.  The analysis includes the experimental uncertainties and the
uncertainties from the detector cross sections and from the minor
fluxes.  When the data are fit, the $pp$ and CNO fluxes are changed
freely for each $\phi(\mbox{Be})$ and $\phi(\mbox{B})$, subject only
to the luminosity constraint.  This plane shown in
Figures~\ref{fig_fff-each} and \ref{fig_fff-comb} essentially
represents every possible astrophysical solution: from the minimum
rate model ($\phi(\mbox{Be}) =
\phi(\mbox{B}) = 0$) to the SSM ($\phi(\mbox{Be}) = \phi(\mbox{B}) =
1$).  That is, any claims --- in the past, present, and future --- of
the astrophysical solutions are represented in
Figures~\ref{fig_fff-each} and \ref{fig_fff-comb} as long as they
satisfy the minimal assumptions.

When each experiment is fit separately, the Kamiokande data constrain
the $^8$B flux, and the Homestake and gallium experiments constrain
both the $^7$Be and $^8$B fluxes; constraints at the 90\% C.L.\ from
each experiment are shown in Fig.~\ref{fig_fff-each}.  In this model
$\phi(\mbox{Be}) = \phi(\mbox{B}) = 0$ is completely consistent with
the Homestake and gallium observation, since we can always adjust the
$pp$ and CNO fluxes to the observed rates.  As seen in
Fig.~\ref{fig_fff-each}, each of the experiments is inconsistent with
the SSM prediction.

When the observations are fit combined, astrophysical solutions in
general are excluded, and the argument comes in three steps.

The $\chi^2$ value of the joint fit is very poor.  In fact if
(nonphysical) negative fluxes were allowed, the $\chi^2$ minimum is
obtained for $\phi(\mbox{Be})/\phi(\mbox{Be})_{\mbox{\scriptsize SSM}}
\sim -0.5$.  Within the physical region, the best fit is obtained for
$\phi(\mbox{Be})/\phi(\mbox{Be})_{\mbox{\scriptsize SSM}} < 0.08$ and
$\phi(\mbox{B})/\phi(\mbox{B})_{\mbox{\scriptsize SSM}} = 0.37 \pm
0.04$ (1$\sigma$), but only with $\chi^2 = 4.1$ for 1 d.f.
Statistically this fit is excluded at 96\% C.L.  In other words, any
astrophysical solutions consistent with our minimal assumptions are
excluded at 96\% C.L. by the observations.

Secondly, even if I allow the 4\% chance and accept the solution, the
obtained fluxes do not physically make sense at all.  Since $^8$B is
produced by the reaction $p + {}^7\mbox{Be} \rightarrow {}^8\mbox{B} +
\gamma$, essentially any reduction in the $^7$Be flux should cause at
least an equal reduction in the $^8$B flux, and additional temperature
sensitivity reduces the $^8$B flux more.  Therefore, unless the
uncertainty of the $^7$Be electron capture rate is grossly
underestimated or there is some independent mechanism to suppress only
the $^7$Be neutrinos, any realistic model should be in the region
$\phi(\mbox{Be})/\phi(\mbox{Be})_{\mbox{\scriptsize SSM}}
\geq \phi(\mbox{B})/\phi(\mbox{B})_{\mbox{\scriptsize SSM}}$ in
Fig.~\ref{fig_fff-comb}, contrary to the observations.

Finally various standard and nonstandard solar models are plotted in
Fig.~\ref{fig_fff-comb}.  The Bahcall-Pinsonneault SSM with 90\%
C.L. uncertainty \cite{Bahcall-Pinsonneault}, the Bahcall-Ulrich 1000
Monte Carlo SSMs \cite{Bahcall-Ulrich}, the Turck-Chi\`eze--Lopes SSM
\cite{Turck-Chieze-Lopes}, the solar models with opacity reduced
by 10 and 20\% \cite{Dearborn}, the large $S_{11}$ models by
Castellani et al.\ \cite{Castellani-Degl'Innocenti-Fiorentini}, the
low Z model by Bahcall and Ulrich \cite{Bahcall-Ulrich,Bahcall-book},
the WIMP model by Gilliland et al.\ \cite{WIMPs,Bahcall-book}, and the
Dar-Shaviv model \cite{Dar-Shaviv}.  Also shown are the power law
dependence of the fluxes to $T_C$ and $S_{11}$ obtained by Bahcall and
Ulrich \cite{Bahcall-Ulrich,Bahcall-book}.  None of those models is
even close to the observations.

This complete failure of our model forces us to conclude that at least
one of the original assumptions is wrong.  We have repeated the
analysis without the luminosity constraint;
\footnote{
The only motivation for this is to allow for the possibility of models
with extremely rapid variation in the solar core (i.e., faster than
the 10$^4$ year time scale for photons to diffuse to the surface),
such as the model of Ref.~[\citenum{De-Rujula-Glashow}].
}
the exclusion of the model is somewhat weaken (the exclusion is at
95\% C.L.), but the conclusion remains essentially the same.
Therefore we have to conclude that either of the other two assumptions
is incorrect: (a) there is a distortion of the $^8$B flux, that is,
new neutrino physics, {\it or} (b) one or more of the experiments is
wrong: in particular, either the Homestake or Kamiokande data must be
grossly wrong.
\footnote{
It was argued by Davis \cite{Davis,Davis-Erice} that the $^8$B flux
observed in Homestake and Kamiokande is consistent, and it is
explained by the mixing model, which assumes an {\it ad hoc} mixture
in the solar interior \cite{Mixing-model}.  The comparison is,
however, made with highly selective data sets: the data only during
1986 -- 1990 are used.  Furthermore the Kamiokande II rate is taken
from the data set with the energy threshold 9.3 MeV, ignoring the data
with the threshold 7.5 MeV.  The data used in the comparison is about
15\% of the total Homestake data and 30\% of the currently available
data from Kamiokande II and III.  Given the consistency among the
Kamiokande results with different thresholds
\cite{Kamiokande-II,Kamiokande-III}, this selection of the Kamiokande
data is unjustified.  The selection of the data during a particular
time period is justified only when a time variation in the flux is
assumed.  The mixing model, to which the selected data were compared,
does not predict time variations, and therefore the theory should be
compared to the averaged data.  The mixing model (with a slow mixing
for 40\% mass) predicts the Kamiokande, Homestake, and gallium rates
to be 0.44 BP-SSM, 3.6 SNU, and 104 SNU , while the averaged data
(without any selection) are 0.51 $\pm$ 0.07 BP-SSM, 2.32 $\pm$ 0.26
SNU, and 78 $\pm$ 10 SNU, respectively.  This mixing model is
inconsistent with the observations.  As discussed by Merryfield
\cite{Merryfield}, the mixing model is also disfavored by the
helioseismology data.
}


\subsection{	What if Homestake data were ignored?  }

As discussed in the previous section, the most compelling motivation
for the particle physics solutions such as the MSW effect comes from
the comparison of the Homestake and Kamiokande data, and, if one of
the two were ignored, the astrophysical solutions cannot be dismissed
outright.  Some argue that we would lose motivations for new neutrino
physics if the Homestake data were ignored.  This, however, is
misleading: at present we have no realistic solar model that are
consistent both with the Kamiokande and gallium results within their
uncertainties.

\begin{figure}[p]
\postscript{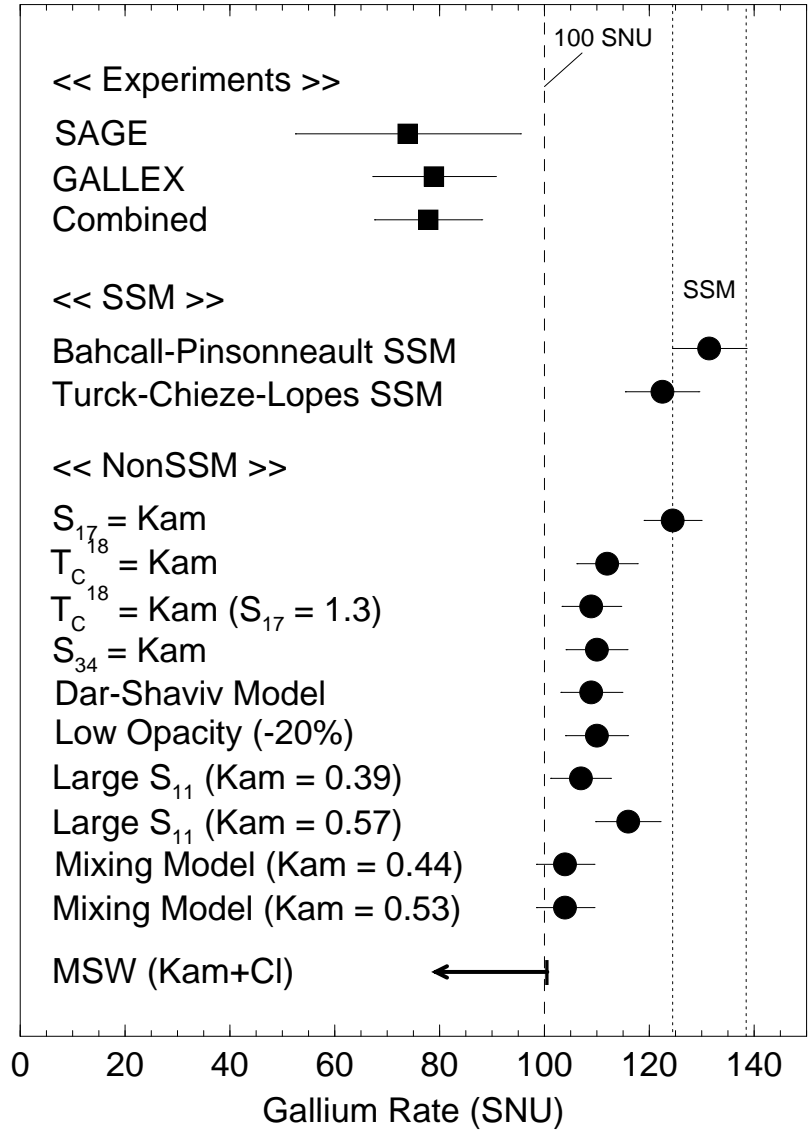}{0.70\hsize}                               
\caption{
%
%
The rate of the gallium experiments, the gallium rates of the SSMs,
and of the nonstandard solar models which predict the $^8$B flux
consistent with or close to the flux observed in Kamiokande.  See text
for details.  The nonstandard solar models consistent with Kamiokande
predict the gallium rate $R_{\mbox{\protect\scriptsize Ga}} > $ 100
SNU.  The MSW solution obtained from the combined Kamiokande and
Homestake data predicts $R_{\mbox{\protect\scriptsize Ga}} < $ 100
SNU, consistent with the data.
}
\label{fig_ga-rate}
\end{figure}

I address the issue by considering the solar models with the $^8$B
flux consistent with the observed Kamiokande rate
\cite{Shi-Schramm-Dearborn}.  The Monte Carlo
SSMs with such a constraint were discussed in
Ref.~[\citenum{Bahcall-Bethe}], but I also allow nonstandard solar
models.  The neutrino flux varies significantly from model to model in
nonstandard solar models in general, but the fact that about half of
the SSM $^8$B flux is seen strongly constrains the nonstandard solar
models, and yields predictions for the gallium rate and the
helioseismology observations.  I list in Fig.~\ref{fig_ga-rate} the
expected gallium rates from various nonstandard models that are
consistent with or close to the $^8$B flux observed by Kamiokande: the
model in which $S_{17}$ is adjusted from the Bahcall-Pinsonneault
model to the observed Kamiokande rate ($0.51
\pm 0.07$ SSM), the low $T_C$ model, the low $T_C$ model with $S_{17}$
increased by 30\%, the model with $S_{34}$ reduced to the Kamiokande
rate, the Dar-Shaviv model
\cite{Dar-Shaviv}, the low opacity model ($-$20\%)
\cite{Dearborn}, the model with $S_{11}$ increased by 30\%
\cite{Castellani-Degl'Innocenti-Fiorentini} and 25\%
\cite{Turck-Chieze-Lopes}, and the mixing models by Sienkiewicz et al
\cite{Mixing-model}.  The uncertainties include the uncertainty of the
$^8$B flux due to the Kamiokande uncertainty (14\%), but the dominant
contribution is from the gallium cross section uncertainty.  From this
list, I obtain
\begin{equation}
   \mbox{Gallium rate consistent with Kamiokande} \geq 100 \; \mbox{SNU},
\end{equation}
while the combined gallium rate of SAGE and GALLEX is 78 $\pm$ 10 SNU.

Obviously it is still premature to rule out those nonstandard solar
models given the current large experimental uncertainties.  But the
observations and the theoretical predictions are showing
discrepancies.  Also other solar observations might exclude those
models.  The helioseismology observations provide information on the
sound speed throughout the radiative and convective zone (but not in
the core region where the theoretical and experimental uncertainties
are still large).  As discussed by other speakers at this conference,
the observations give significant constraints in solar model building:
the SSM is consistent with the helioseismology data at the 1\% level,
and any large departures from the SSM will be likely to destroy this
agreement.  In fact, some of the nonstandard solar models listed in
Fig.~\ref{fig_fff-comb} and also other nonstandard models such as the
low $T_C$ model \cite{Gough}, the large $S_{11}$ model
\cite{Turck-Chieze-Lopes}, the mixing model \cite{Merryfield}, and
the low Y model \cite{Gough} are in conflict with the helioseismology
observations.

While waiting for the gallium detectors to be calibrated and the
statistics of the gallium data to improve, this line of theoretical
investigations should be encouraged.  I hope that the possibility of
solar models consistent with the Kamiokande and the gallium data will
be pursued, and the list of nonstandard models as in
Fig.~\ref{fig_fff-comb} will be extended to obtain the minimum rate
expected for the gallium experiments.  Also, each and every model should
be confronted with the helioseismology observations.  If one could
successfully find a model consistent with the Kamiokande and gallium
as well as the helioseismology observations, it could be the solution
to the solar neutrino problem: the SSM and the Homestake experiment
might be wrong after all.  If, however, one could not find such a
model after entirely discarding the Homestake data, our motivation for
considering particle physics solutions will be complete.


\section{		MSW Solutions			}
\label{sec_msw}

There are many proposed particle physics solutions to the solar
neutrino problem: the MSW effect with two or more flavors, vacuum
oscillations, a large neutrino magnetic moment, neutrino decays,
violations of the equivalence principle, etc.  I discuss the two state
MSW oscillations \cite{MSW}, since it requires only the minimal
extension of the standard model (mass and mixing of neutrinos) without
contradicting with any other existing observational constraints, and
also without requiring any unnatural fine-tunings of the parameters.
Furthermore, many extensions of the standard electroweak model with
the see-saw mechanism \cite{see-saw-mechanism} suggest neutrino mass
consistent with the MSW parameter range
\cite{Langacker-Neutrino-Telescopes}, providing a theoretical
motivation to look for neutrino mass in the solar neutrino
experiments.  I also consider the MSW oscillations for sterile
neutrinos and the MSW solutions in the context of nonstandard solar
models.  The MSW analysis of the solar neutrino data was carried out
by many authors
\cite{Bahcall-Haxton,Shi-Schramm,Gelb-Kwong-Rosen,Krastev-Petcov,%
Krauss-Gates-White,Fogli-Lisi-Montanino,Fiorentini-etal}, but here I
follow the recent analysis of
Ref.~[\citenum{BHKL,HL-Earth,HL-MSW-analysis}], which includes the
theoretical uncertainties and their correlations, the Earth effect,
the Kamiokande II day-night data, and the improved definition of the
confidence levels.


\subsection{		Earth Effect			}

One interesting, but not-well publicized, aspect of the MSW mechanism
is the Earth effect \cite{Earth-effect} on the solar neutrinos.  The
electron neutrinos produced in the Sun transform to another species of
neutrino by the MSW effect in the Sun, but, for a certain parameter
range, they can transform back to $\nu_e$'s during the night due to
the matter effect as the neutrinos propagate through the Earth.  Such
an effect can be seen as a difference in the observed rate between day
and night, or, more drastically, as signal variations as a function of
time during the night since the oscillation probabilities depend
sensitively on the neutrino path length in the Earth and also on the
densities that the neutrino go through (i.e., the mantle only or both
the mantle and the core).  The Earth effect is only relevant for a
limited region of $\Delta m^2$ and $\sin^2 2\theta$, but, if measured,
it would be an indisputable evidence for the MSW solution.  Given the
high statistics in the next generation experiments, the MSW parameters
($\Delta m^2$ and $\sin^22\theta$) could be determined precisely by
measuring time and energy dependence.  Since the measurement depends
only on the time (and energy) dependence, but not on the absolute
neutrino flux, the determination of $\Delta m^2$ and $\sin^22\theta$
would be free from the solar model uncertainties and also from the
time-independent systematic uncertainties in the experiments; the
uncertainties from the Earth density profile is negligibly small.

The MSW transition is most significant when the resonance condition
is satisfied:
\begin{equation}
\label{eqn_MSW-resonance}
	{\Delta m^2 \over 2E } \cos2\theta = \sqrt{2}\, G_F \, n,
\end{equation}
where $E$ is the neutrino energy, $G_F$ is the Fermi constant; $n =
n_e$ for oscillations to $\nu_\mu$ or $\nu_\tau$, and $n = n_e - n_n /
2$ for oscillations to sterile neutrinos, where $n_e$ and $n_n$ are
the electron and neutron densities.  The density profile of the Earth
is approximately described as two step functions: the core (10 -- 12.5
g/cm$^3$) and the mantle (4 -- 5 g/cm$^3$).  A more detailed density
profile can be found in Ref.~[\citenum{Earth-profile}].  For solar
neutrinos energies, the parameter space relevant for the Earth effect
is $\Delta m^2 \sim 10^{-7} - 10^{-5} \; \mbox{eV}\; ^2$ and
$\sin^22\theta \geq 0.01$.  To compare the MSW predictions to the
averaged rates of observations, one needs to average the $\nu_e$
survival probability over the night, since it depends on the path
length and also on the density profile through which the neutrinos
propagate.  The survival probability also vary each night due to the
obliquity of the Earth rotation axis and should be averaged over a
year.  Also it differs for each experiment due to the detector
latitudes.  This averaging process requires intensive computations.

\begin{figure}[phbt]
\begin{tabular}{c}
\setlength{\epsfxsize}{\mswsize}
\subfigure[Day-time rate]{\epsfbox{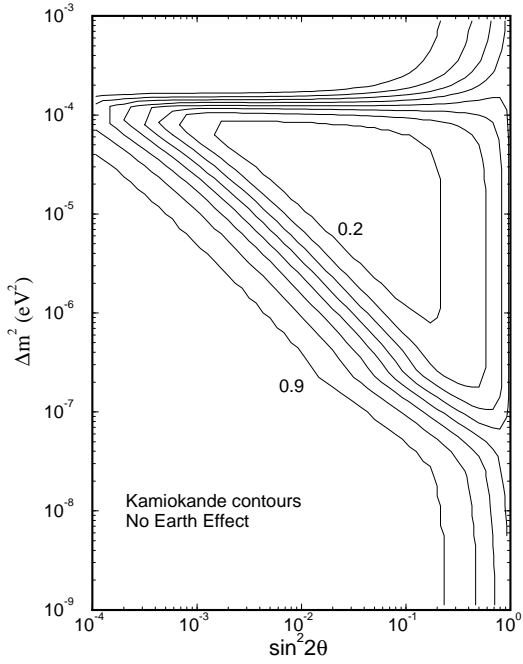}}
\hspace{3em}
\subfigure[Night-time rate]{\epsfbox{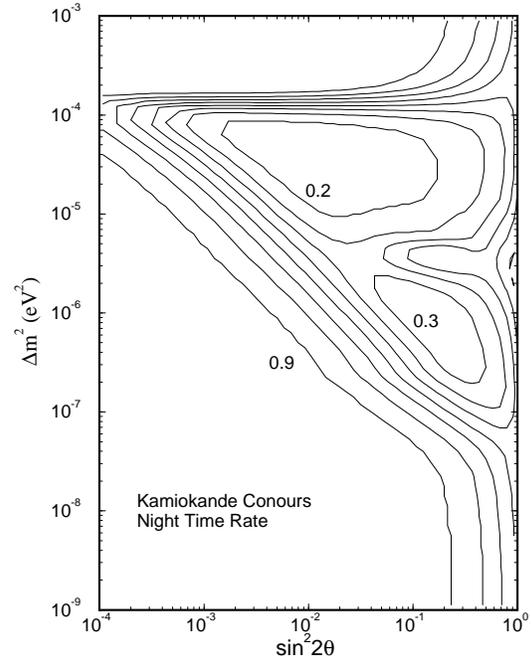}}  \\
\setlength{\epsfxsize}{\mswsize}
\subfigure[Kam II day-night data]{\epsfbox{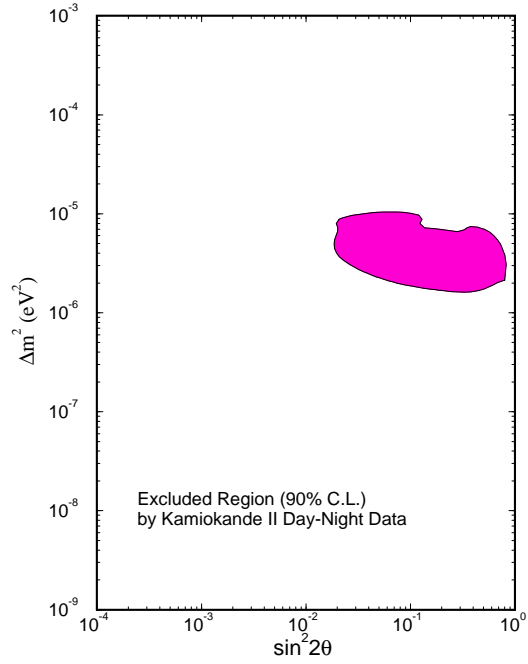}}
\end{tabular}
\caption{
%
%
The contours of the signal-to-SSM ratio in Kamiokande for (a) day time
and (b) night time.  (c) Also shown is the region excluded by the
Kamiokande II day-night data with 6 time bins.
}
\label{fig_c_kam}
\end{figure}

The contours of the signal-to-SSM ratio for the Kamiokande experiment
are shown in Fig.~\ref{fig_c_kam} for the day-time rate (a) and the
night-time rate (b).  For a certain parameter space, the night-time
enhancement is huge: for example, the night-time rate is almost three
times the day-time rate for $\Delta m^2 \sim 3 \times 10^{-6}\;
\mbox{eV}\, ^{2}$ and $\sin^22\theta \sim 0.2$.

Real-time counting experiments such as Kamiokande are capable of
observing the day-night effect.  In fact the Kamiokande II
collaboration has published the day-night data with 6 bins (one
day-time data point and 5 night-time data points for different angles
between the Sun direction and the nadir at the detector; the binning
corresponds to selecting neutrinos with different path lengths in the
Earth).  The data show no enhancement of the signal at night, albeit
still large statistical uncertainties, and exclude the parameter space
shown in Fig.~\ref{fig_c_kam} (c).  This exclusion of parameter space
is insensitive to the solar model uncertainties and independent of
solar models.

The Earth effect has significant phenomenological implications to the
existing and future direct-counting experiments.  The day-night data
are available so far only from the Kamiokande II experiment.  The
additional data from Kamiokande III is expected to double the
statistics, and the updated analysis should further constrain the
parameter space.  In particular, a significant day-night effect is
expected in the large-angle solution of the combined observations,
which might be observable if the Kamiokande II data and the (still
unpublished) Kamiokande III day-night data are combined.  With large
counting rates in the future solar neutrino experiments, one should be
able to observe a day-night effect at a few per cent level, and the
parameter space in which the contours in Fig.~\ref{fig_c_kam} (b) are
distorted from Fig.~\ref{fig_c_kam} (a) should produce observable
signals.  If the day-night effect is observed, its time and energy
dependence provides a clean determination of $\Delta m^2$ and
$\sin^22\theta$ without referring to solar model calculations.  The
detailed predictions of the day-night effect for SNO,
Super-Kamiokande, and Kamiokande obtained from the MSW solutions of
the combined observations will be given in
Section~\ref{sec_future-exp}.


\subsection{		MSW Analysis			}

Given a solar model, either standard or nonstandard, the MSW mechanism
is a solid, calculable theory.  It can be quantitatively tested and
constrained, and the following is the questions to be addressed in the
MSW analysis:
\begin{enumerate}
\item	Is the MSW hypothesis acceptable by the observations?

\item	If acceptable, what is the parameter space allowed by the
	observations?

\item	What are the predictions for the future experiments from the
	obtained parameters?
\end{enumerate}
The first question tests the whole idea of the MSW theory, including
the assumption of neutrino mass and mixing.  The answer to the second
question will provide important information for the neutrino mass
spectrum and mixing matrix of the lepton sector.  And most
importantly, the MSW theory is verifiable or falsifiable by the next
generation experiments.

To answer those questions quantitatively, it is important to include
relevant theoretical uncertainties in the analysis.  Especially the
$^8$B flux uncertainty in the Bahcall-Pinsonneault SSM is 14\% and
comparable to the experimental uncertainties of Homestake (11\%) and
Kamiokande (14\%).  In the Turck-Chi\`eze --Lopes SSM, the $^8$B flux
uncertainty is 25\% and dominates the experimental uncertainties.

Equally important, but often ignored, are the correlations among the
theoretical uncertainties.  The leading uncertainty from the $^8$B
flux is strongly correlated among the experiments, especially between
Homestake and Kamiokande.  Also the uncertainties are correlated
between different flux components.  If, for example, the opacity were
lower (or the core temperature were lower), both the $^7$Be flux and
$^8$B flux would be reduced.  Therefore a careful estimation of the
theoretical uncertainties and their correlations are essential in
determining the neutrino parameters.  In fact if the theoretical
uncertainties were ignored, there would be no allowed solution in the
large-angle region at 90\% C.L.  If the correlations were ignored, a
significantly larger region would be obtained for the large-angle
solution \cite{HL-MSW-analysis}.

\begin{figure}[htb]
\postscript{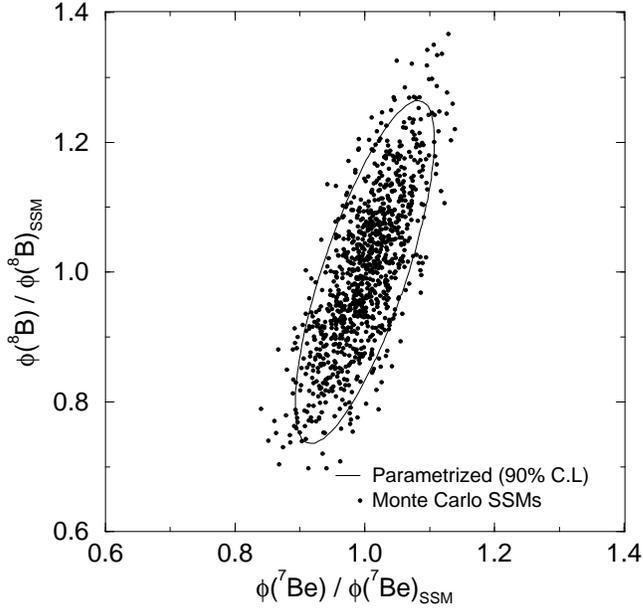}{0.60\hsize}                          
\caption{
%
%
The $^7$Be and $^8$B flux distributions of the Bahcall-Ulrich 1000
Monte Carlo SSMs (dots).  Also shown (solid line) are the flux
uncertainties (at 90\% C.L.) parametrized by the central temperature
and the relevant nuclear cross sections.  The two methods provide
consistent flux uncertainty estimations.  Taken from
Ref.~[\protect\citenum{HL-MSW-analysis}].
}
\label{fig_BeB}
\end{figure}

We have parameterized the SSM flux uncertainties by the uncertainties
in the central temperature and in the relevant nuclear reaction cross
sections ($S_{17}$ and $S_{34}$) \cite{HL-MSW-analysis,BHKL,HL-Earth}.
It was shown explicitly that our treatment reproduces the flux
uncertainties and correlations obtained from the Bahcall-Ulrich 1000
Monte Carlo SSMs \cite{HL-MSW-analysis}.  The parameterization method
can be easily generalized to other solar models, for which the Monte
Carlo estimations are unavailable.  The $^7$Be and $^8$B flux
distributions of the Bahcall-Ulrich 1000 Monte Carlo SSMs and our
parametrized uncertainty estimation are displayed in
Fig.~\ref{fig_BeB}.  The amplitudes and the correlation of the flux
uncertainties are consistent between the two methods.

One ambiguity in determining the neutrino parameters is the definition
of confidence levels.  The general prescription in this case is that,
once the MSW mechanism is accepted as a viable solution, the
probability density of the parameter space allowed by the observations
is distributed throughout the $\log\Delta m^2 - \log\sin^22\theta$ plane, and
is given as $P = N \exp ( -\chi^2/2 )$, where $N$ is the normalization
factor and the $\chi^2$ is calculated for each combination of $\Delta
m^2$ and $\sin^22\theta$
\cite{HL-MSW-analysis}.  When the probability distribution is
approximated as a gaussian around the $\chi^2$ minimum, the allowed
region is determined by $\chi^2(\Delta m^2, \sin^22\theta) \leq
\chi^2_{\mbox{\scriptsize min}} + \Delta
\chi^2$ with $\Delta \chi^2$ = 4.6, 6.0, and 9.2 for 90, 95,
and 99\% C.L., respectively.  The $\Delta \chi^2$ is determined by the
fact that the probability density is distributed in the 2-dimensional
parameter space, and does not depend on the number of data points
which enter the $\chi^2$ calculation.  The assumption of one gaussian
distribution is questionable for the MSW analysis, which yields
multiple $\chi^2$ minima in the parameter space.  We have generalized
the $\Delta \chi^2$ method in the presence of multiple $\chi^2$ minima
\cite{HL-MSW-analysis}, and will use the improved definition in the
joint analysis.

We use the Petcov formula \cite{Petcov,Kuo-Pantaleone} to calculate
the $\nu_e$ survival probability, and have compared the results to
other analytic approximations (the Parke formula with the Landau-Zener
approximation \cite{Parke} and the Pizzochero formula
\cite{Pizzochero}), and found that the difference in the combined fit
is completely negligible.  Also our calculations were compared to
those of Fiorentini et al.\ \cite{Fiorentini-etal}, who solve the
neutrino propagation equation numerically: the agreement was
excellent.

Our MSW calculations include the MSW double-crossing, the neutrino
production distributions in the Sun \cite{Bahcall-Pinsonneault}, the
electron density distribution \cite{Bahcall-Pinsonneault}, the
detector cross sections \cite{Bahcall-Ulrich,Bahcall-book}, and the
energy resolution and the detector efficiency for Kamiokande
\cite{Kamiokande-II}.

The uncertainties from the neutrino production distributions and from
the electron density distributions are studied by comparing three
different solar models (but using the same flux magnitude)
\cite{HL-MSW-analysis}: the Bahcall-Ulrich SSM and the
Bahcall-Pinsonneault SSM with and without the helium diffusion effect.
No difference was observed.  We have also changed by $\pm$10\% each
the peak location of the production distribution, the electron density
scale height, and the core-mantle boundary in the Earth, which enters
into the Earth effect calculation; the combined solutions were
insensitive to the changes \cite{HL-MSW-analysis}.


\subsection{		Updated Results			}

\begin{figure}[phbt]
\begin{center}
\begin{tabular}{c}
\setlength{\epsfxsize}{\mswsize}
\subfigure[Combined results]{\epsfbox{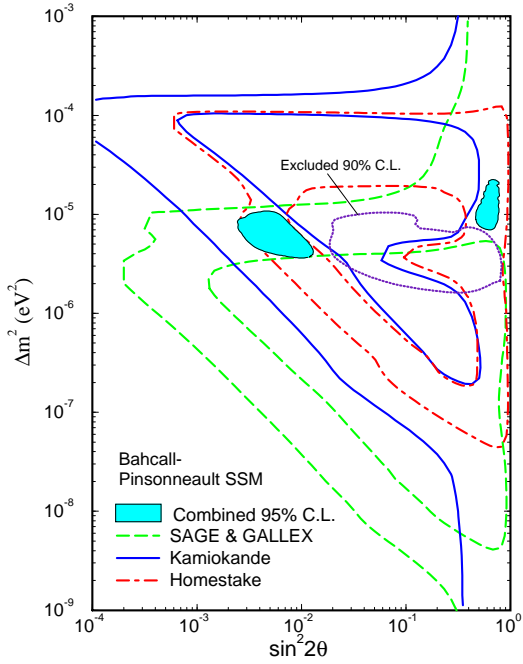}}
\hspace{3em}
\subfigure[90, 95, and 99\%  C.L.]{\epsfbox{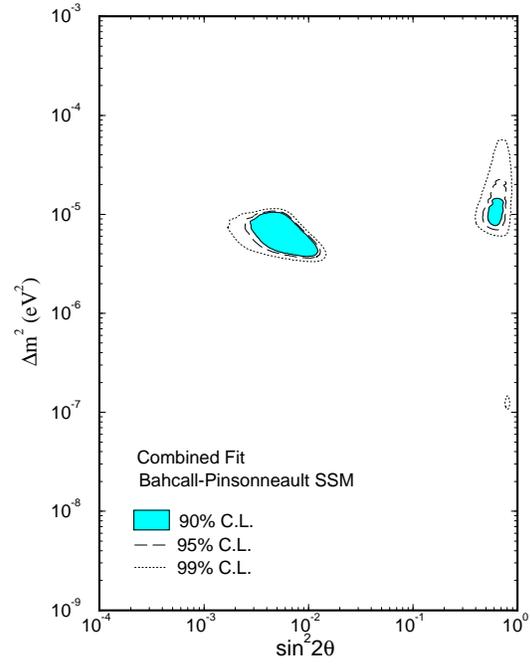}}
\end{tabular}
\caption{
%
%
(a) The MSW solutions allowed at 95\% C.L.\ by the combined
observations of the Homestake, the combined gallium SAGE and GALLEX,
the Kamiokande II including six day-night data points, and the
Kamiokande III data.  Also shown are the regions allowed by each of
the experiments and the region excluded by the Kamiokande II day-night
data (dotted line).  (b) The combined allowed regions at 90, 95, and
99\% C.L.  This is an updated analysis of
Ref.~[\protect\citenum{HL-MSW-analysis}].
}
\label{fig_MSW-comb}
\end{center}
\end{figure}

The updated MSW results including the most recent GALLEX, SAGE, and
Kamiokande III data are presented in Fig.~\ref{fig_MSW-comb} (a).  The
joint fit includes the Homestake, combined gallium, Kamiokande II
day-night data (6 data points) and the Kamiokande III data.  We have
incorporated the Earth effect, the theoretical and experimental
uncertainties and their correlations.  The Bahcall-Pinsonneault SSM
with helium diffusion effect is assumed unless otherwise mentioned.
The improved confidence level definition is robust under the
assumption that the probability density is distributed throughout the
parameter space and the distribution is described as a multiple gaussian
distribution, i.e., the allowed regions should have an elliptic shape,
which is a good approximation in our case.  Fig.~\ref{fig_MSW-comb}
(b) shows the combined allowed regions for 90, 95, and 99\% C.L.  Note
that there is a third allowed region for $\Delta m^2 \sim 1.2 \times
10^{-7} \, \mbox{eV}\, ^2$ and $\sin^22\theta \sim 0.8$, but only at
99\% C.L.; this region is disfavored by the Kamiokande II day-night
data and also by the combined result of GALLEX and SAGE.

\begin{table}
\begin{center}
%
%
\caption{The best fit parameters of the combined MSW solutions.}
\label{tab_MSW-comb}
\vspace{1ex}
\begin{tabular}{ l c c c }
\hline
\hline
		  & Nonadiabatic & Large Angle I & Large Angle II \\
\hline
  $\sin^22\theta$    & $6.8\times 10^{-3}$ & 0.62      & 0.80          \\
  $\Delta m^2$ (eV$^2$)& $6.2\times 10^{-6}$ & $9.7\times 10^{-6}$
                                                   & $1.3\times 10^{-7}$\\
\hline
  $\chi^2$ (7 d.f.) &  2.4              & 6.9		& 12.7             \\
  G.O.F.(\%)        &  94               & 44		& 8              \\
  $P_{\mbox{\scriptsize relative}}$ (\%)
                  &  92.2     	      & 7.5             & 0.3              \\
\hline
\hline
\end{tabular}
\end{center}
\end{table}

The best fit parameters, $\chi^2$ values, the goodness-of-fit (G.O.F),
and relative probabilities are listed for each solution in
Table~\ref{tab_MSW-comb}.  G.O.F is the probability of obtaining by
chance a $\chi^2$ value equal to or larger than the obtained $\chi^2$.
The relative probability is the probability of finding the true
solution in this region when the probability density is assumed to be
a gaussian distribution for each solution
\cite{HL-MSW-analysis}.

Note that the small-angle (nonadiabatic) solution provides an
excellent fit: the MSW solution is an acceptable theory by the
combined observations.  In fact $\chi^2 = 2.4 $ for 7 d.f. (9 data
points -- 2 parameters) is too good.  This is due to a small
contribution from the Kamiokande II 6 day-night data point ($\chi^2
\sim 2.1$), and without the day-night data, the $\chi^2/$1 d.f. is 0.3
(58\% C.L.), which is reasonable.

\begin{figure}[hbt]
\postscript{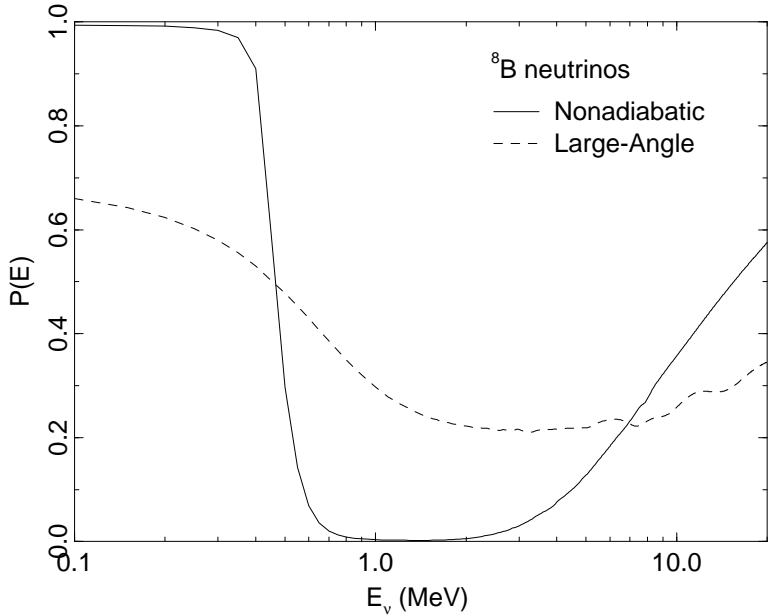}{0.65\hsize}                       
\caption{
%
%
The energy dependence of the two MSW solutions allowed by the combined
observations.  The small-angle (nonadiabatic) solution gives a better
fit than the large-angle solution.  The survival probability depends
on the neutrino production distribution, and the plot assumes the
distribution of the $^8$B neutrinos, which is concentrated around the
center of the Sun.  The survival probability for the two $^7$Be lines
is essentially the same; for the $pp$ neutrinos, the production region
is more wide spread, and consequently the decrease of the survival
probability around $E \sim 0.5$ MeV is less steep.
}
\label{fig_PE}
\end{figure}

Compared to the small-angle (nonadiabatic) solution, the two
large-angle solutions are disfavored by the observations.  This is due
to the characteristic energy dependence of the solutions shown in
Fig.~\ref{fig_PE}.  For the small-angle (nonadiabatic) solution, the
$^7$Be neutrinos and the lower energy part of the $^8$B neutrinos are
suppressed, consistent with the observed relative rates of Homestake
and Kamiokande.  On the other hand, the energy dependence for the
large angle solution is mild, and the central values of Homestake and
Kamiokande cannot be simultaneously given by the MSW effect.  This is
similar to the reason why the astrophysical solutions are disfavored,
but is less severe: in the large-angle MSW case, the converted
$\nu_\mu$ or $\nu_\tau$ can still interact in the Kamiokande detector
via the neutral current interaction (with about 1/6 the cross
section), reducing the inferred $\nu_e$ survival probability to a
value close to that of Homestake.  There is also a slight enhancement
of the survival probability at larger energies due to the Earth
effect.

\begin{figure}[phbt]
\postscript{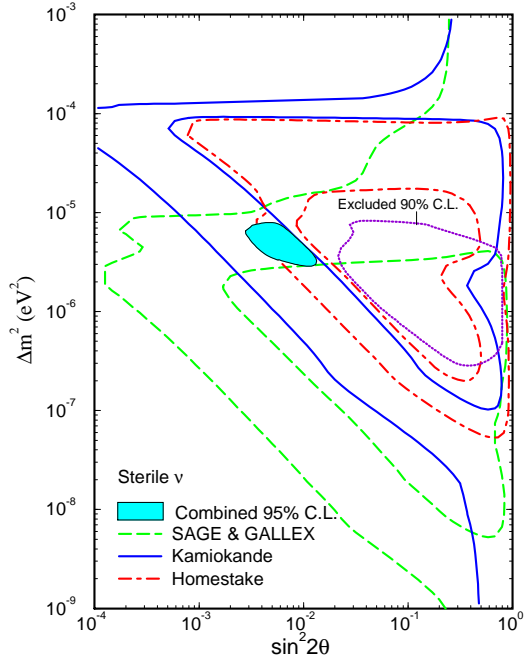}{\mswsize}        
\caption{
%
%
The parameter space allowed for oscillations for sterile neutrinos.
There is no allowed regions for large angles even at 99\% C.L., which
is consistent with the big-bang nucleosynthesis constraint.  Taken
from Ref.~[\protect\citenum{HL-MSW-analysis}].
}
\label{fig_MSW-sterile}
\end{figure}

The MSW oscillations also can be considered for sterile neutrinos, and
the allowed parameter space is displayed in
Fig.~\ref{fig_MSW-sterile}.  There is no allowed large-angle region
even at 99\% C.L., since the absence of the neutral current events in
Kamiokande
\footnote{
The MSW conversion formula is also modified for sterile neutrinos by
the replacement $n_e \rightarrow n_e - n_n/2$ (see
Eqn.~\ref{eqn_MSW-resonance}), but the effect on the combined
solutions are numerically negligible. }
makes the $\nu_e$ survival probabilities effectively larger than for
flavor oscillations.  The large-angle regions are therefore strongly
disfavored, much like the astrophysical solutions.  The exclusion of
the large-angle region for sterile neutrinos has been independently
obtained from neutrino counting in big-bang nucleosynthesis
\cite{sterile}.

\begin{figure}[pbth]
\postscript{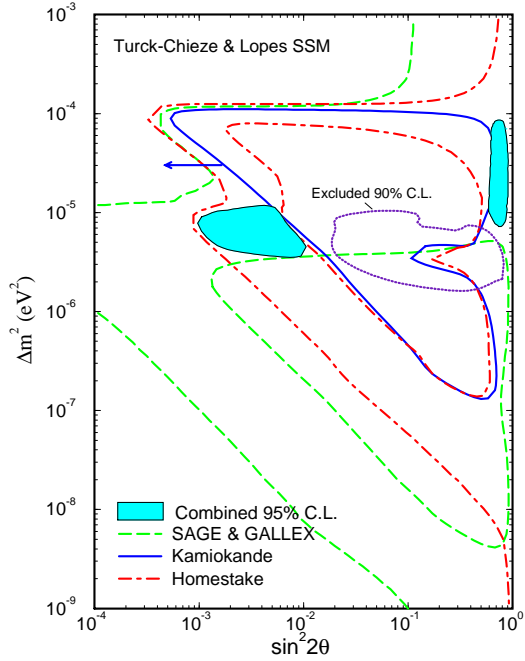}{\mswsize}                     
\caption{
%
%
The allowed parameter space when the Turck-Chi\`eze--Lopes SSM is
assumed.  Taken from Ref.~[\protect\citenum{HL-MSW-analysis}].
}
\label{fig_MSW-TCL}
\end{figure}

The MSW regions for the Turck-Chi\`eze--Lopes model is displayed in
Fig.~\ref{fig_MSW-TCL}.  This model does not include the particle
diffusion effect, and also the neutrino production distributions and
the electron density distributions have not been published.  The
calculation is done by using those of the Bahcall-Pinsonneault model.
The obtained allowed regions are slightly shifted outward of the MSW
triangle compared to the Bahcall-Pinsonneault solutions due to the
smaller predicted fluxes.  Also the allowed regions are noticeably
larger, reflecting that the flux uncertainties are more conservative
in the Turck-Chi\`eze--Lopes SSM.  Therefore the correct uncertainty
estimation is important in constraining neutrino parameters, and
underestimation of the flux uncertainties can lead to underestimation
of the uncertainties of $\Delta m^2$ and $\sin^22\theta$.

\subsection{		MSW in Nonstandard Solar Models }

\begin{figure}[phbt]
\postscript{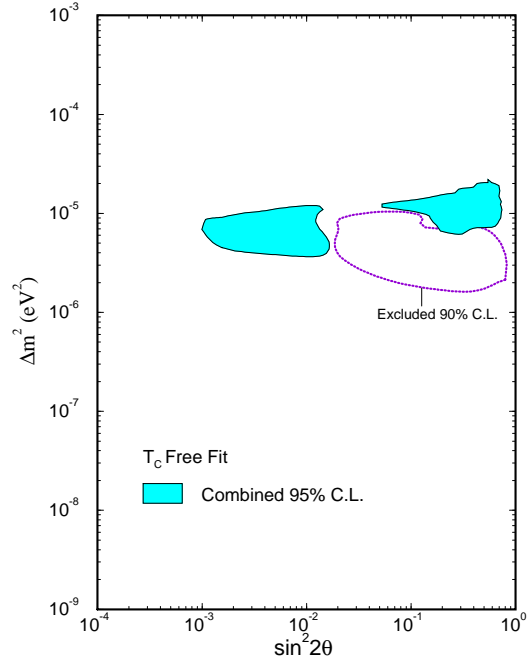}{\mswsize}                    
\caption{
%
%
The allowed parameter space when the central temperature is allowed to
change freely.  The data constrain $T_C = 1.02 \pm 0.02 \; (1 \sigma)$.
Taken from Ref.~[\protect\citenum{HL-MSW-analysis}].
}
\label{fig_MSW-Tcfree}
\end{figure}

The SSM is assumed in the MSW analysis discussed in the previous
section.  Even if the MSW is proven, however, that does not confirm
that the SSM is correct; the MSW effect can also take place in the
nonstandard solar models.  The SSM is still needs to be tested and
calibrated experimentally.  To address such a possibility, we have
carried out the MSW analysis while the core temperature is allowed to
change freely.  The neutrino fluxes are changed according to the power
law obtained by Bahcall and Ulrich (Eqn.~\ref{eqn_Tc-power-law}),
except that the $T_C$ dependence of the pp flux is obtained from the
luminosity constraint (Eqn.~\ref{eqn_luminosity}).  From a three
parameter fit ($\Delta m^2$, $\sin^22\theta$, and $T_C$), we
constrained
\begin{equation}
		T_C = 1.02 \pm 0.02 \hspace{0.5em} (1 \sigma),
\end{equation}
with $\chi^2_{\mbox{\scriptsize min}} = 2.1$ for 6 degrees of freedom
(d.f.), where $T_C$ is in units of the Bahcall-Pinsonneault SSM ($15.57
\times 10^6 \; \mbox{K}$).  Interestingly the existing observations
constrain the central temperature of the Sun at the 2\% level even in
the presence of the MSW effect.  (Note that there is no consistent
$T_C$ to describe the observations without the MSW effect.  See
Section~\ref{sec_cool-sun-model}.)  Also the value obtained for $T_C$
is consistent with the Bahcall-Pinsonneault SSM value $T_C = 1 \pm
0.006$.  If the MSW is proven, the above will be the first
observational constraint on the core temperature of the Sun at the 2\%
level.  The combined MSW regions with free $T_C$ are shown in
Fig.~\ref{fig_MSW-Tcfree}.

\begin{figure}[phbt]
\postscript{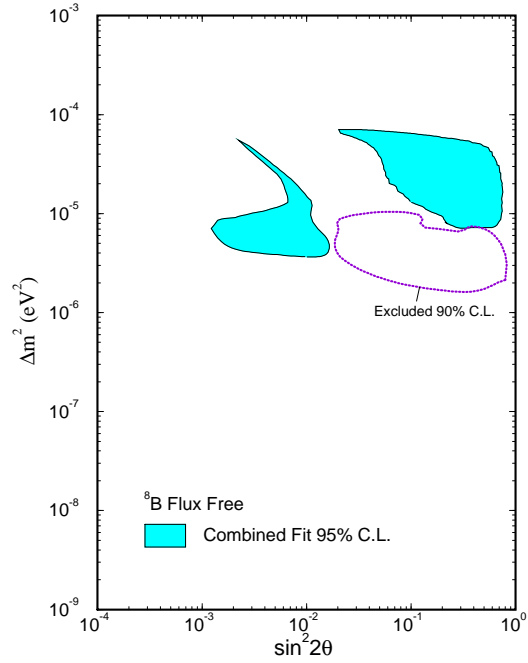}{\mswsize}                    
\caption{
%
%
The allowed parameter space when the $^8$B flux is used as a free
parameter.  Taken from Ref.~[\protect\citenum{HL-MSW-analysis}].
}
\label{fig_MSW-B8free}
\end{figure}

Since the $^8$B flux has the largest uncertainty (14\% in the
Bahcall-Pinsonneault SSM and 25\% in the Turck-Chi\`eze--Lopes model),
we have also carried out the analysis with the $^8$B flux as a free
parameter.  The constraint obtained from the observations is weak,
but consistent with the SSM:
\begin{equation}
        \phi(\mbox{B}) / \phi(\mbox{B})_{\mbox{\scriptsize SSM}}
        = 1.43 + 0.65 - 0.42 \; (1 \sigma).
\end{equation}
The allowed regions for this case is shown in Fig.~\ref{fig_MSW-B8free}.

\begin{figure}[p]
\begin{center}
\begin{tabular}{c}
\setlength{\epsfxsize}{\mswsize}
\subfigure[Low opacity]{\epsfbox{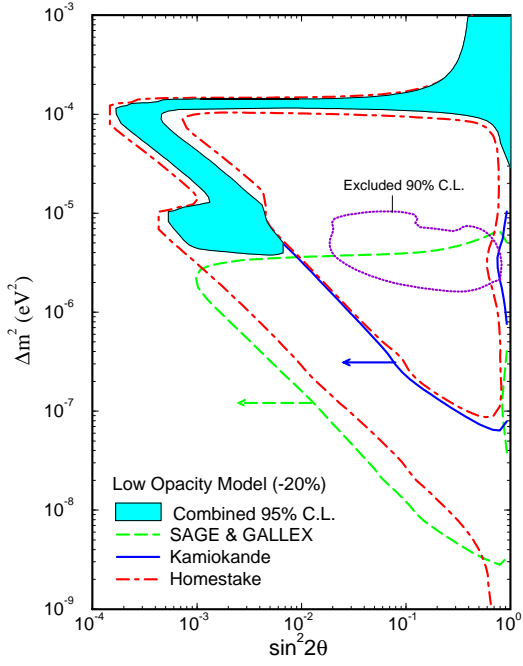}}
\hspace{3em}
\subfigure[Large $S_{11}$]{\epsfbox{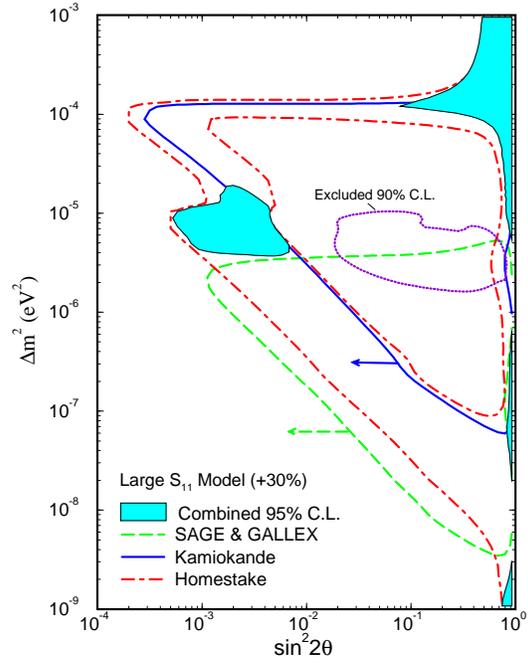}}  \\
\setlength{\epsfxsize}{\mswsize}
\subfigure[High Y model]{\epsfbox{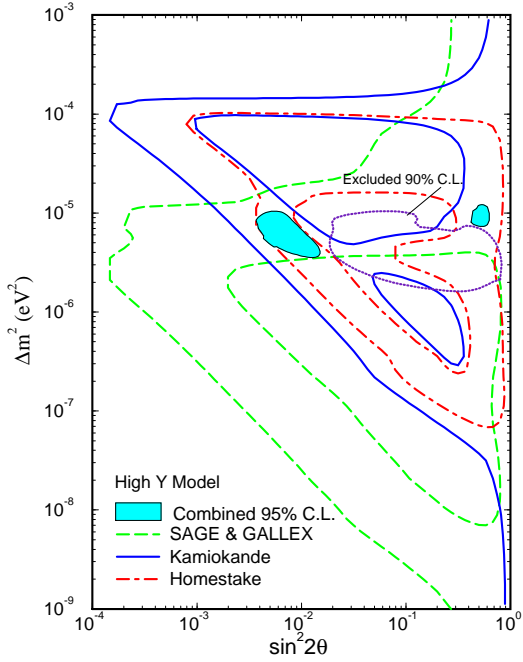}}
\hspace{3em}
\subfigure[Maximum rate model]{\epsfbox{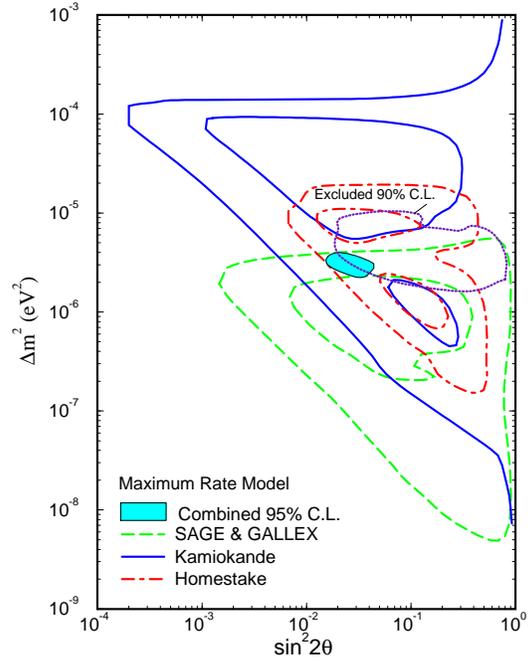}}  \\
\end{tabular}
\end{center}
\vspace{-4ex}
\caption{
%
%
The allowed parameter space for nonstandard solar models: (a) the low
opacity model and (b) the large $S_{11}$ model predict smaller
neutrino fluxes than the SSM, while (c) the high Y model and (d) the
maximum rate model predict larger fluxes.  Taken from
Ref.~[\protect\citenum{HL-MSW-analysis}].
}
\label{fig_MSW-nonssms}
\end{figure}

We have also considered the MSW solutions for nonstandard solar models
that are explicitly constructed, and the results are shown for four
nonstandard solar models in Fig.~\ref{fig_MSW-nonssms}.  Two of them
predict smaller initial fluxes compared to the SSM: the low opacity
model \cite{Dearborn} and the large $S_{11}$ model
\cite{Castellani-Degl'Innocenti-Fiorentini}.  The other two predict
larger fluxes than the SSM: the high Y model \cite{Bahcall-Ulrich} and
the maximum rate model \cite{Bahcall-Pinsonneault}.  With nonstandard
input parameters or nonstandard core temperatures, a wider MSW
parameter space is allowed.

Among the above nonstandard solar models, the maximum rate model and
the large $S_{11}$ model are significantly different from the SSM, and
are disfavored by the solar neutrino data even with the MSW effect.
But nonstandard core temperature within $T_C \sim 1 - 1.04$ is
perfectly consistent with the observation.  A further constraint on
$T_C$ (or the neutrino flux) from the helioseismology observations is
welcome.


\subsection{		Observational Constraints and Hints for
			Neutrino Mass and Mixing }

\begin{figure}[p]
\postscript{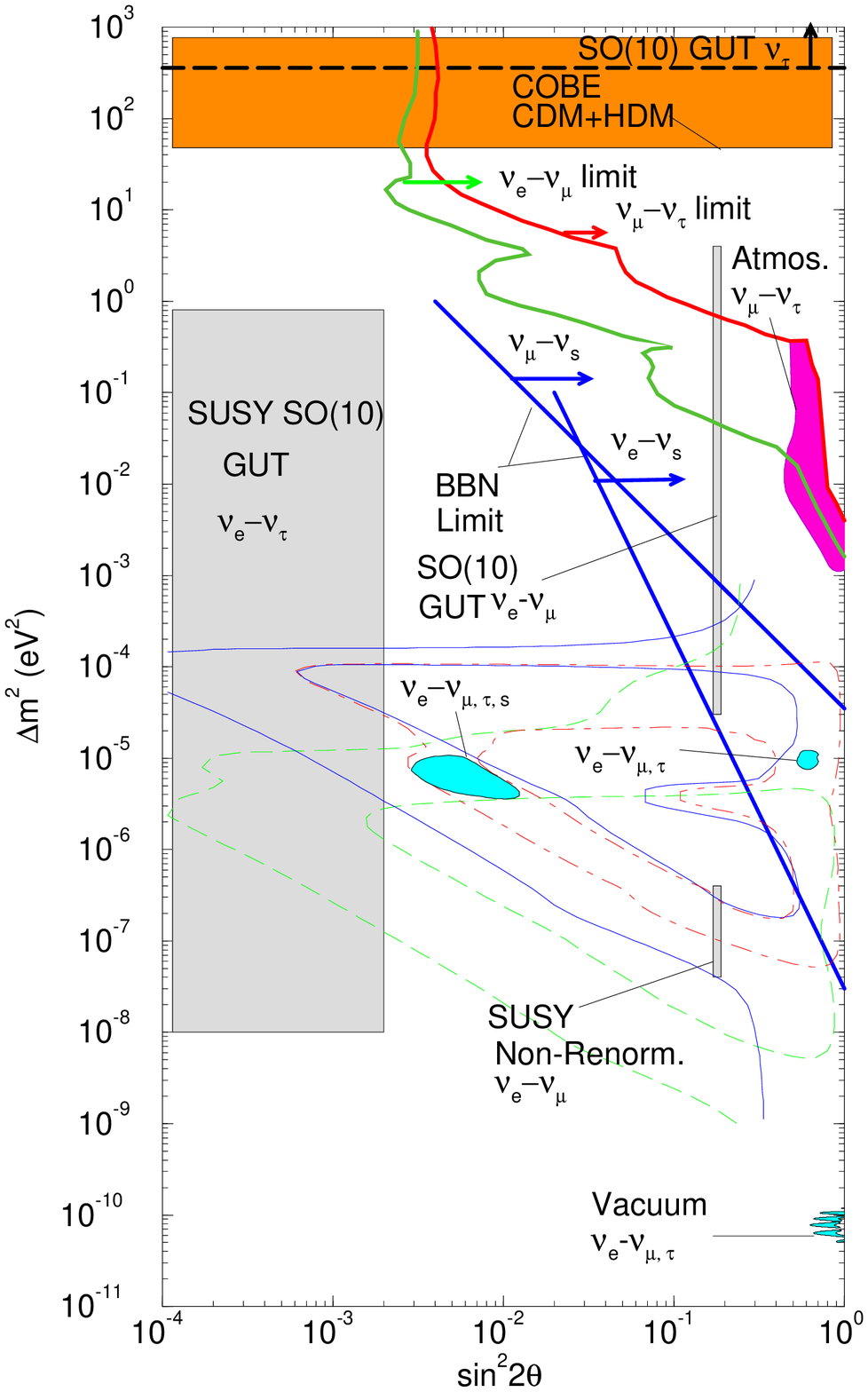}{0.90\hsize}                     
\vspace{4ex}
\caption{
%
%
Various observational constraints and hints for neutrino mass and
mixing.  Also shown are various theoretical predictions. See text for
details.
}
\label{fig_big-picture}
\end{figure}

There are two observational hints of neutrino mass other than solar
neutrinos: the oscillation interpretation of the atmospheric neutrino
deficit \cite{atmospheric-neutrino-problem}, and the cold plus hot
dark matter scenario for the cosmic background radiation measurements
and the large-scale structure observations \cite{CPHDM}.  The
parameter ranges for those hypotheses are displayed in
Fig.~\ref{fig_big-picture}, along with the MSW and vacuum oscillation
solutions \cite{NH-vacuum} of the solar neutrino problem.  The
constraints from the oscillation experiments
\cite{oscillation-limit} and the constraints for oscillations to sterile
neutrinos from big-bang nucleosynthesis \cite{Shi-Schramm-Fields} are
also shown.

Fig.~\ref{fig_big-picture} also shows the parameter space of various
theoretical predictions involving the see-saw mechanism
\cite{see-saw-mechanism}: the $\nu_e - \nu_\mu$ mixing in the
$SO_{10}$ grand unified theory (GUT) with an intermediate scale
breaking \cite{BKL,Langacker-Neutrino-Telescopes}, the $\nu_e -
\nu_\tau$ mixing in the supersymmetric $SO_{10}$ GUT
\cite{BKL,Langacker-Neutrino-Telescopes}, and the $\nu_e - \nu_\mu$
mixing in the superstring-inspired model with nonrenormalizable
operators \cite{Cvetic-Langacker} (there is no compelling prediction
for the mixing angles for this model).  The mixing angles shown for
these theories (Fig.~\ref{fig_big-picture}) are the corresponding
quark mixing angles ($V_{\mbox{\scriptsize lepton}} \sim
V_{\mbox{\scriptsize CKM}}$).  In the $SO_{10}$ GUT, the $\nu_\tau$
mass is expected to be in a range relevant to the cosmological hot
dark matter.  The detailed mass range and mixing angles are model
dependent, and can be different from what is shown.  The general idea
of grand unification and the see-saw mechanism suggests the neutrino
mass scales shown in Fig.~\ref{fig_big-picture}, and there is a solid
motivation to look for neutrino mass in the MSW range relevant for
solar neutrinos.

The next question is whether this picture is reality, or just a
fantasy of particle physicists, and, more importantly, whether we can
prove or disprove such theoretical ideas.  For solar neutrinos,
experimental quests are directly aiming at the detection of neutrino
oscillations, and the next generation experiments of the Sudbury
Neutrino Observatory (SNO), Super-Kamiokande, BOREXINO, and ICARUS
should either confirm or rule out the neutrino oscillation hypothesis.
For other neutrino mass ranges, two accelerator-based oscillation
experiments at CERN (CHORUS and NOMAD) will start operating in the
spring of 1994, and explore the cosmologically interesting $\nu_\tau$
mass range down to small mixing angles ($\Delta m^2 > 1 - 10 \;
\mbox{eV}\, ^2$ and $\sin^22\theta < 2
\times 10^{-4}$ for $\nu_\mu \leftrightarrow \nu_\tau$ oscillations).
The proposed long-baseline experiments of Brookhaven and
Fermilab-SOUDAN will explore the parameter space suggested by the
atmospheric neutrino deficit.

\subsection{		MSW Predictions for Future Experiments }
\label{sec_future-exp}

As discussed in the previous sections, we have compelling reasons to
consider the MSW mechanism as a solution to the solar neutrino
problem: a lack of viable astrophysical explanations provides a strong
motivation for particle physics solutions, while the data are
consistent with the MSW descriptions; the idea of GUTs and the see-saw
mechanism suggests a neutrino mass range consistent with the MSW
solutions.

It is still premature, however, to convince ourselves that neutrino
mass and mixing are the cause of the solar neutrino deficit.  More
data are needed, and the next generation experiments such as SNO
\cite{SNO}, Super-Kamiokande\cite{Super-Kamiokande}, BOREXINO
\cite{BOREXINO}, and ICARUS \cite{ICARUS} should provide a
definitive answer for neutrino oscillation hypothesis.  There are two
theoretical questions which should be answered by the new experiments.
\begin{itemize}
\item	Calibrate the solar model.  The neutrino flux should be
	determined component by component from experimental data.

\item	Determine the solution and the parameter space.  In particular
	distinguish the MSW solutions from astrophysical solutions
	and also from other particle physics solutions, and constrain
	$\Delta m^2$ and $\sin^22\theta$.
\end{itemize}
The procedure of doing these can be complicated, since the the
determination of the neutrino parameters depends on the neutrino flux
and, in turn, the determination of the neutrino flux requires a
knowledge of the neutrino parameters.  These can be accomplished by
the next generation experiments with high counting rates and various
reaction modes.  The reaction modes of the experiments with projected
counting rates are summarized in Table~\ref{tab_future-exp}.

\begin{table}[hbt]
\begin{center}
\caption{The future solar neutrino experiments.  CC, NC, and ES stand
for charged current, neutral current, and elastic scattering mode.}
\vspace{1ex}
\label{tab_future-exp}
\scriptsize
\begin{tabular}{l  c l c}
\hline \hline \\
Experiment & Commission date &\hspace{2em} Reaction mode & Event rate  \\[2ex]
\hline\\
SNO		& Fall, 1995	& CC: $\nu_e+d\rightarrow p+p+e^-$
				& 3,000 events/yr                      \\[2ex]
		&		& NC: $ \nu_{e,\mu,\tau} + d
				      \rightarrow \nu_{e, \mu, \tau} + p + n$
		                &     3,000 events/yr   		\\[2ex]
		&		& ES: $ \nu_{e,\mu,\tau} + e^- \rightarrow
				      \nu_{e,\mu,\tau} + e^- $
				& 200 events/yr   		   \\[2ex]
\hline\\
Super-Kamiokande& April, 1996   & ES: $ \nu_{e,\mu,\tau} + e^- \rightarrow
				      \nu_{e,\mu,\tau} + e^- $
				& 8,000 events/yr   		   \\[2ex]
\hline\\
BOREXINO	& Late 1990's   & ES ($^7$Be):
				  $ \nu_{e,\mu,\tau} + e^- \rightarrow
				    \nu_{e,\mu,\tau} + e^- $
				& 18,000 events/yr (SSM) 	   \\[2ex]
\hline\\
ICARUS		& $\sim$ 2000   & ES: $ \nu_{e,\mu,\tau} + e^- \rightarrow
				        \nu_{e,\mu,\tau} + e^- $
				& 2,900 events/yr/module           \\[2ex]
		&		& CC: $ \nu_e + {}^{40}\mbox{Ar} \rightarrow
				      {}^{40}\mbox{K}^*+e^-$
		 	        &  2,400 events/yr/module           \\[2ex]
\hline\\
Iodine		& Late 1990's   & CC: $\nu_e + {}^{127}\mbox{I} \rightarrow
				      {}^{127}\mbox{Xe} + e^-$
				&                                   \\[2ex]
\hline 
\hline 
\end{tabular}
\normalsize
\end{center}
\end{table}

Given the current data and the SSM, one can yield robust predictions
for future experiments to test the MSW hypothesis.  The MSW mechanism
provides a solid, calculable theoretical frame work; the theoretical
uncertainties are, by and large, under control.  The predictions
include the charged to neutral current ratio, the spectrum
distortions, and the day-night difference.  The charged to neutral
current ratio (CC/NC) is the gold-plated measurement in SNO of
confirming neutrino oscillations; however, whether SNO can distinguish
the MSW solution from astrophysical solutions at the $3 \sigma$ level
is solar model dependent; if the $^8$B flux is smaller by 20\% than
that of the Bahcall-Pinsonneault SSM, the MSW solution may not be
established given the projected systematic uncertainty of $\sim$ 20\%.
On the other hand, energy spectrum distortions and the day-night
effect are completely free from the solar model uncertainties and
theoretically clean observables.  By requiring consistency among those
measurements from different experiments, we should be able to
determine the solution and constrain the parameter space with a
sufficient precision.

\subsubsection{		CC/NC Measurement in SNO 		}

\begin{figure}[hbt]
\begin{center}
\begin{tabular}{c}
\setlength{\epsfxsize}{\mswsize}
\subfigure[Bahcall-Pinsonneault SSM]{\epsfbox{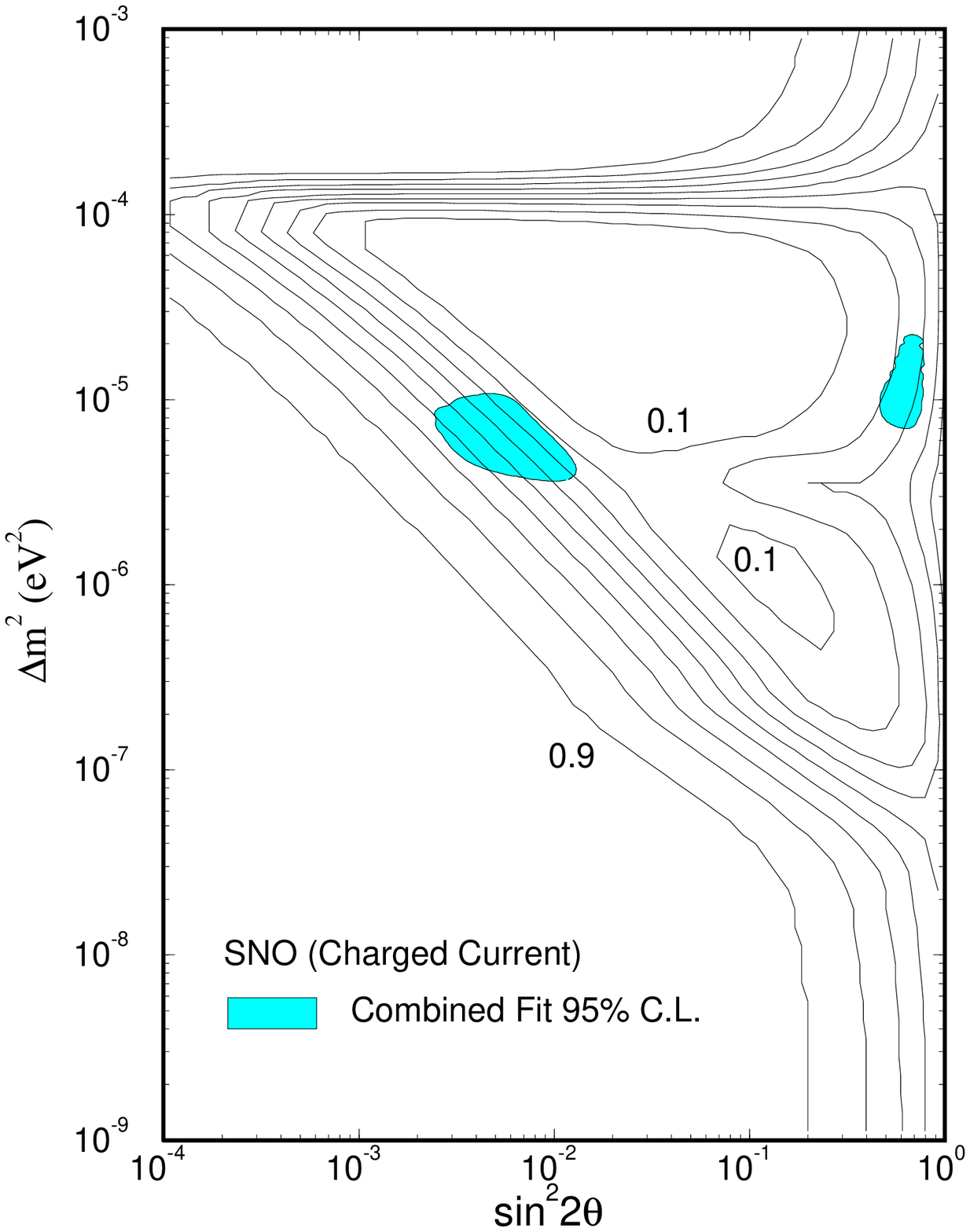}}
\hspace{3em}
\subfigure[Turck-Chi\`eze--Lopes SSM]{\epsfbox{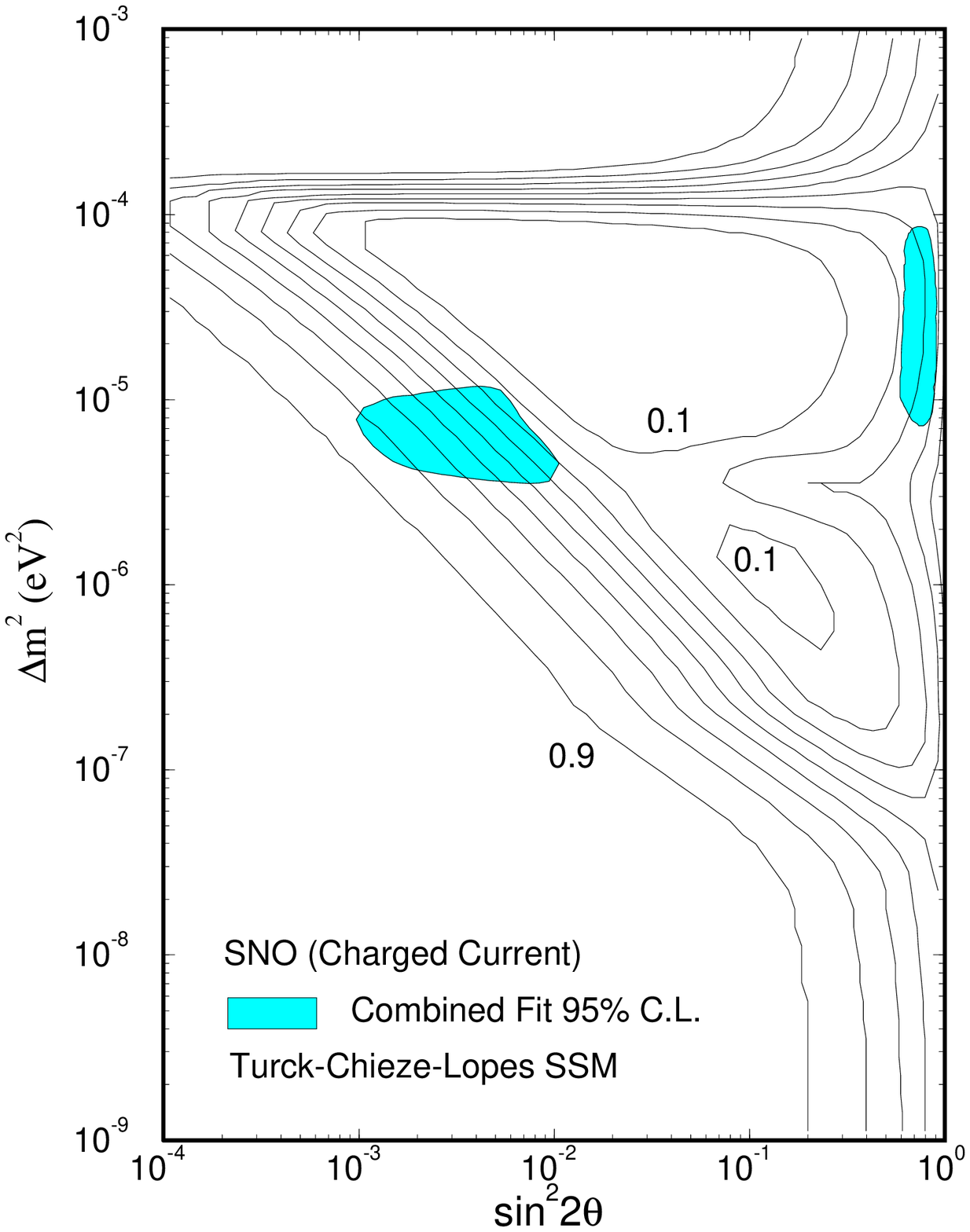}}
\end{tabular}
\end{center}
\caption{
%
%
The contours of the signal-to-SSM ratio for the SNO charged current
measurement with the combined allowed region assuming (a) the
Bahcall-Pinsonneault SSM and (b) the Turck-Chi\`eze--Lopes SSM.  This
is an updated result of Ref.~[\protect\citenum{HL-MSW-analysis}].
}
\label{fig_c_snocc}
\end{figure}

Measuring a depletion of the charged current (CC) rate with respect to
the neutral current (NC) rate in SNO is considered to be the smoking
gun evidence to establish neutrino oscillations.  From the global
solutions assuming the Bahcall-Pinsonneault SSM
[Fig.~\ref{fig_c_snocc} (a)] at 95\% C.L., the ratio (CC/NC) is
expected to be
\begin{equation}
	\frac{\mbox{(CC/NC)}_{\mbox{\scriptsize SNO}} \hspace{0.9em} }
             {\mbox{(CC/NC)}_{\mbox{\scriptsize No Osc}}}  =
		\left\{
		\begin{array}{ l l }
                0.2 - 0.65 & \mbox{ (Nonadiabatic Solution) } \\
		0.2 - 0.3 & \mbox{ (Large-Angle Solution), }
		\end{array}	\right.
\end{equation}
where (CC/NC)$_{\mbox{\scriptsize No Osc}}$ is the ratio expected for
no oscillations or for oscillations into sterile neutrinos.  This
prediction is, however, solar model dependent.  Using the
Turck-Chi\`eze--Lopes model, which predicts a smaller $^8$B flux with
a larger uncertainty, the prediction [Fig.~\ref{fig_c_snocc} (b)]
becomes
\begin{equation}
	\frac{\mbox{(CC/NC)}_{\mbox{\scriptsize SNO}} \hspace{0.9em} }
             {\mbox{(CC/NC)}_{\mbox{\scriptsize No Osc}}} =
		\left\{
		\begin{array}{ l l }
                0.2 - 0.85 & \mbox{ (Nonadiabatic Solution) } \\
		0.2 - 0.4 & \mbox{ (Large-Angle Solution). }
		\end{array}	\right.
\end{equation}
I emphasize that the prediction of whether SNO can distinguish the MSW
solutions from astrophysical solutions by the CC/NC measurement is
strongly dependent on the $^8$B flux and its uncertainty used in
obtaining the current MSW solutions.  With the current projection of
systematic uncertainty at the 20\% level, the distinction with a
3$\sigma$ significance is achievable assuming the Bahcall-Pinsonneault
SSM.  It is no longer guaranteed if the Turck-Chi\`eze--Lopes SSM is
assumed.  Given the large uncertainty in the $S_{17}$ determination, a
$^8$B flux smaller by 20\% than that of the Bahcall-Pinsonneault SSM
is a distinct possibility.  Furthermore, there are additional
uncertainties in other inputs such as opacity calculations.  It is
emphasized that the solar model uncertainties can affect the
predictions for future experiments, and especially the SNO experiment
development can be affected by future improvements in the $S_{17}$
determination.

\subsubsection{		Spectrum Distortions		}

\begin{figure}[p]
\begin{center}
\begin{tabular}{c}
\setlength{\epsfxsize}{0.60\hsize}
\subfigure[SNO (CC)]{\epsfbox{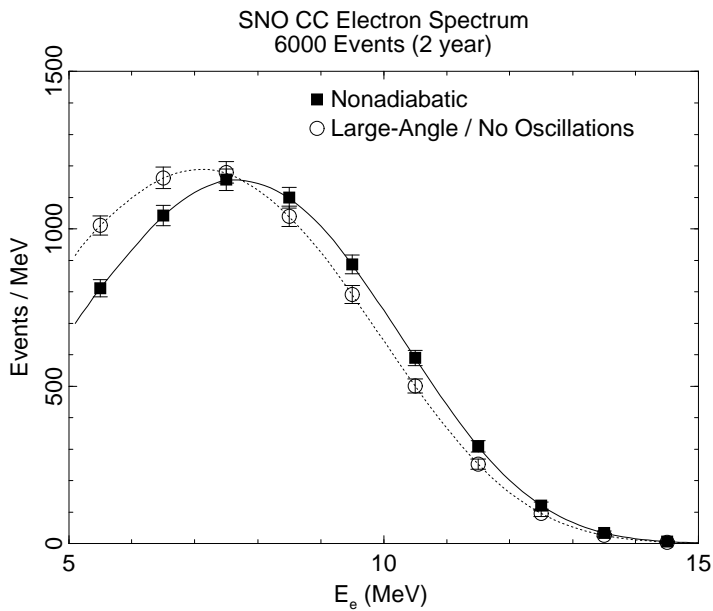}}  \\
\setlength{\epsfxsize}{0.60\hsize}
\subfigure[Super-Kamiokande]{\epsfbox{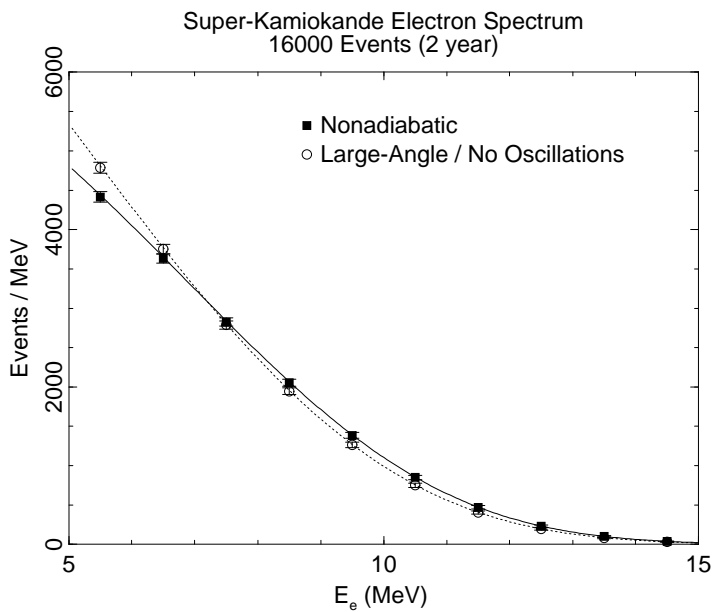}}
\end{tabular}
\end{center}
\vspace{-4ex}
\caption{
%
%
The spectrum shape expected for (a) SNO and (b) Super-Kamiokande.  The
error bars show the statistical uncertainties from totals of 6,000 and
16,000 events, respectively (equivalent to 2 years of operation).  The
large-angle spectrum is normalized to the nonadiabatic spectrum above
5 MeV.  Taken from Ref.~[\protect\citenum{HL-MSW-analysis}].
}
\label{fig_spectrum}
\end{figure}

The $^8$B spectrum measurements in SNO, Super-Kamiokande, and ICARUS
provide a unique opportunity to look for energy dependent effects on
low energy neutrinos with an unprecedented precision.  Astrophysical
effects on the spectrum such as from gravitational red-shifts and
thermal fluctuations are completely negligible at the observable level
\cite{Bahcall-spectrum}, while particle physics effects (or new
neutrino physics) is often energy dependent, and spectrum distortions
are a theoretically clean signal for new physics.  In particular the
MSW small-angle (nonadiabatic) solution predicts spectrum distortions,
suppressing the lower energies of the $^8$B flux more.  The spectrum
shapes predicted by the global MSW solutions are displayed in
Fig.~\ref{fig_spectrum} for SNO (charged current reaction) and
Super-Kamiokande.  The error bars correspond to statistical
uncertainties from 6,000 and 16,000 events, respectively (equivalent
to 2 years of operation).  To compare differences in the shape, the
large-angle spectrum is normalized to the small-angle (nonadiabatic)
spectrum above the threshold (5 MeV).  The calculation includes the
detector energy resolutions and also the differential charged current
cross section
\cite{Nozawa,nu-d-cross-section} for SNO.
\footnote{
It is important in the spectrum shape analysis to include the detector
energy resolutions and the charged current electron spectrum.  The
approximation ``electron energy = neutrino energy -- 1.44 MeV''
significantly overestimates the sensitivity to distortions.  }
The small-angle (nonadiabatic) solution shows a depletion of the
spectrum at lower energies ($<$ 7 MeV), and it is essential to achieve
a low threshold to observe the distortion.  The large-angle solution
shows little distortion, and the shape is identical to astrophysical
solutions (or no oscillation) in the scale of the figure.

\subsubsection{		Day-Night Effect		}

The day-night effect \cite{Earth-effect,HL-Earth} is another purely
particle physics mechanism that is free from astrophysical
uncertainties since the day-night difference (or time dependence) of
the signal is independent of the absolute $^8$B flux.  The
regeneration of $\nu_e$'s in the Earth can be measured not only by the
day-night difference, but also by time-dependence during the night and
energy spectrum distortions.  The time dependence and the spectrum
distortions are often strongly correlated.  The parameter space for
which the day-night effect is most drastic has already been excluded
by the Kamiokande II data, but the next generation experiments with
higher statistics are sensitive to a wider parameter space.  The
day-night effect, if observed, is a dream observable for particle
physicists: the determination of $\Delta m^2$ and $\sin^22\theta$ can
be done precisely by measuring the time dependence and spectrum
distortions, completely independent of solar model predictions.  The
calibrations of the solar neutrino fluxes could then be easily done
once $\Delta m^2$ and $\sin^22\theta$ are determined from the
day-night effect.

\begin{figure}[phtb]
\begin{center}
\begin{tabular}{c}
\setlength{\epsfxsize}{0.45\hsize}
\subfigure[Nonadiabatic]{\epsfbox{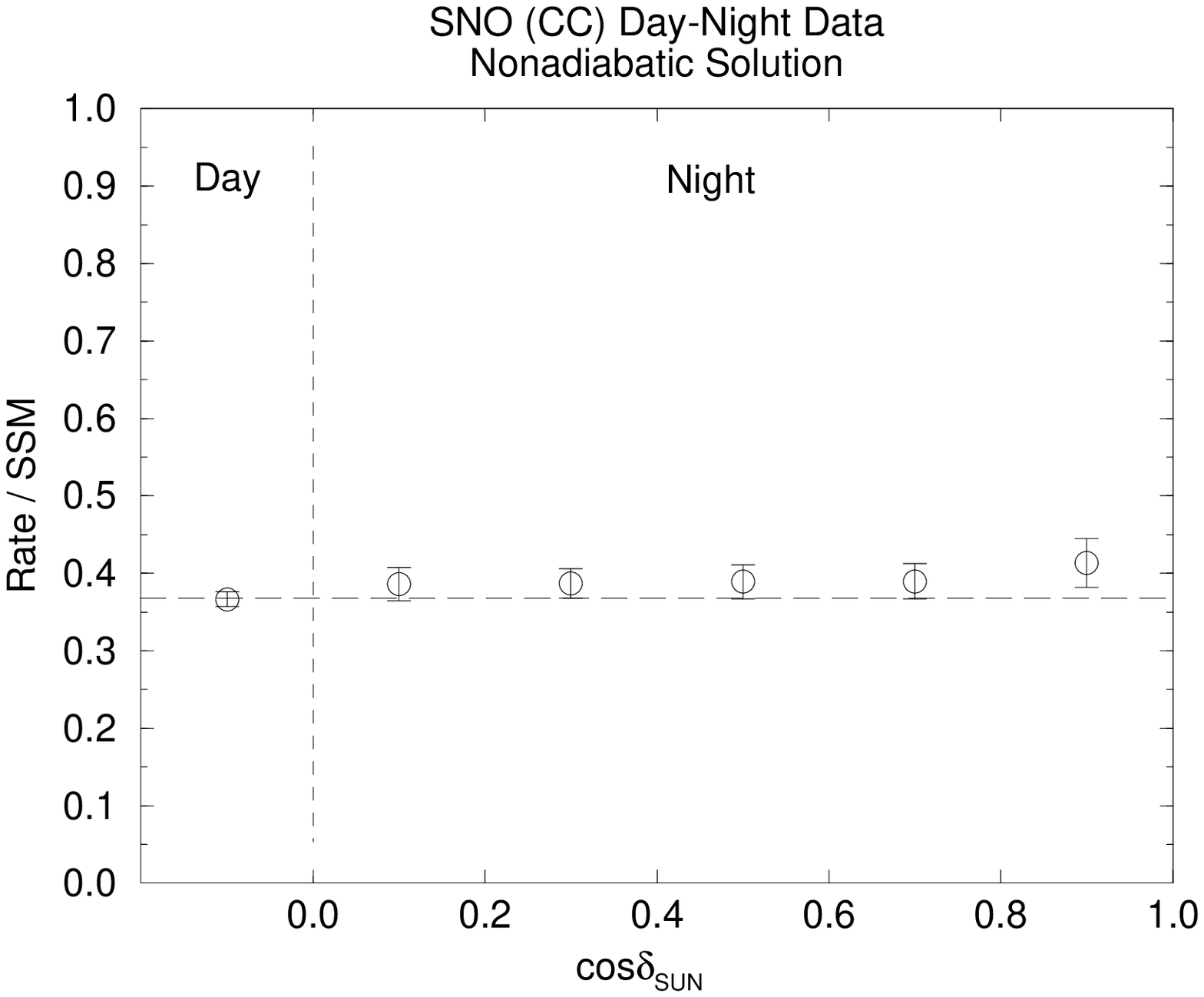}}
\hspace{5mm}
\subfigure[Large-Angle]{\epsfbox{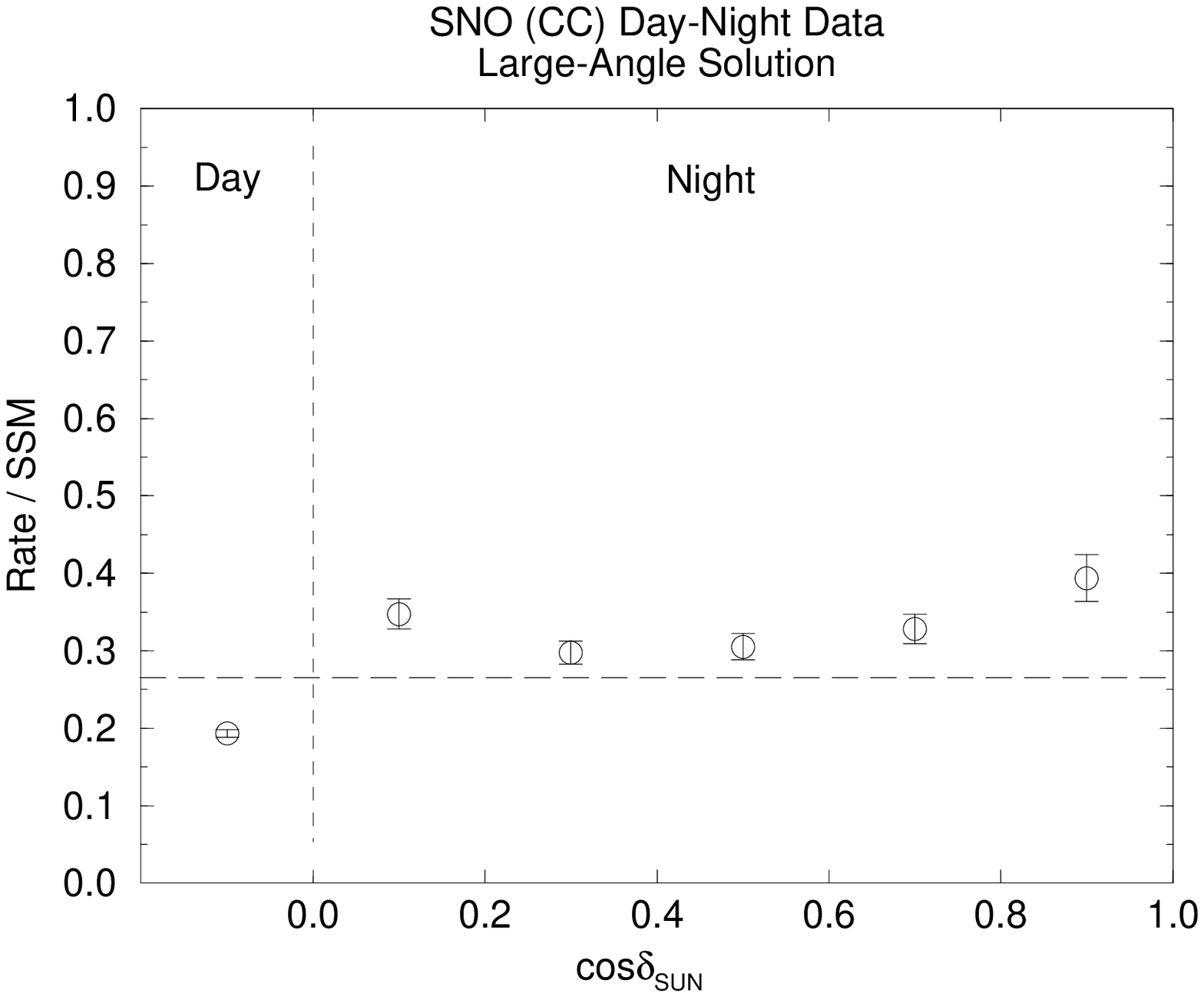}}
\end{tabular}
\end{center}
\caption{
%
%
The day-night effect in SNO for (a) the small-angle (nonadiabatic)
solution and (b) the large-angle solution.
$\delta_{\mbox{\protect\scriptsize sun}}$ is the angle between the
nadir at the detector and the direction in the Sun.  The error bars
are the statistical uncertainties assuming a total of 3,000 events
(equivalent to one year of operation). Taken from
Ref.~[\protect\citenum{HL-MSW-analysis}].
}
\label{fig_dn_sno}
\end{figure}

\begin{figure}[phbt]
\begin{center}
\begin{tabular}{c}
\setlength{\epsfxsize}{0.45\hsize}
\subfigure[Nonadiabatic]{\epsfbox{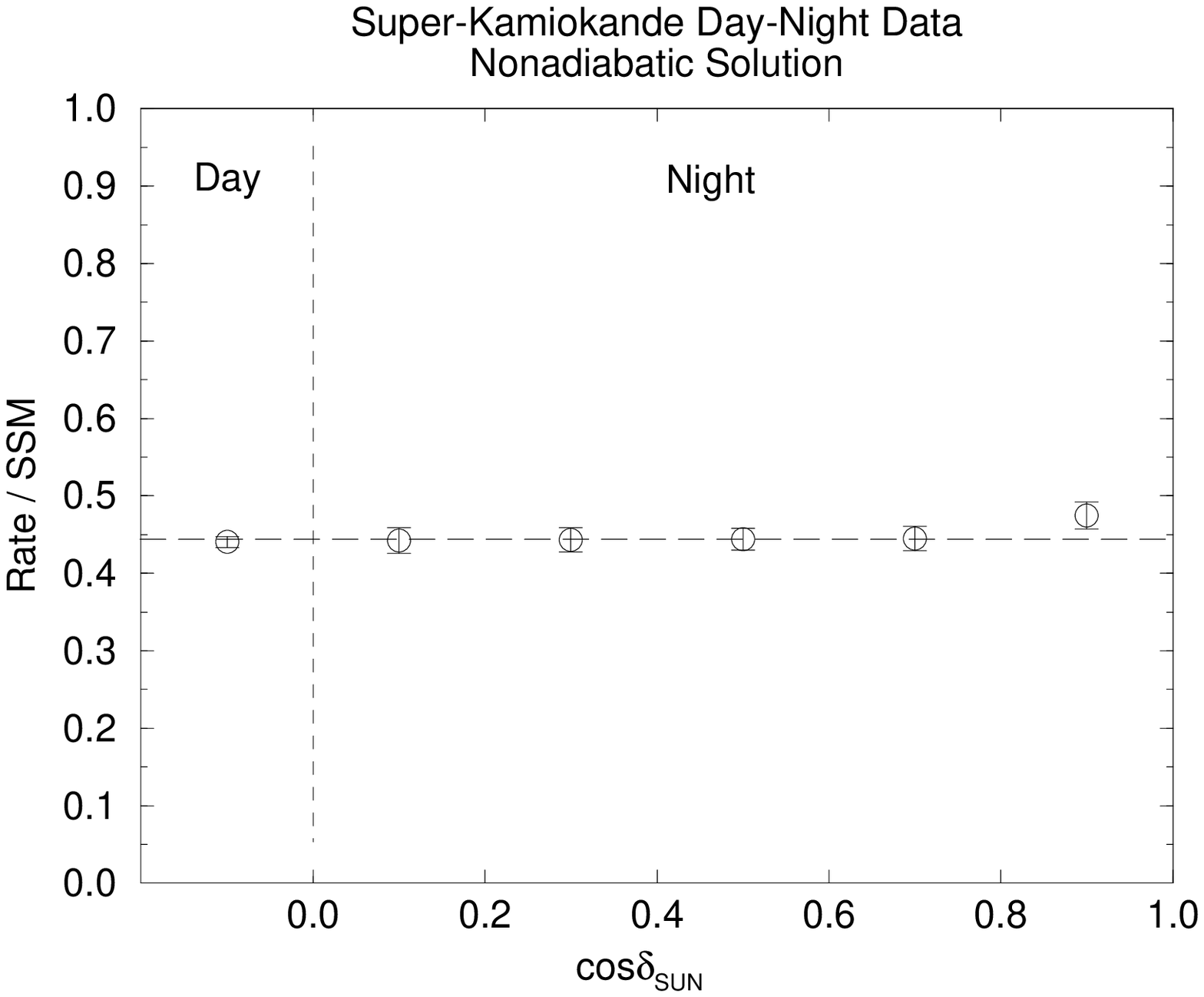}}
\hspace{5mm}
\subfigure[Large-Angle]{\epsfbox{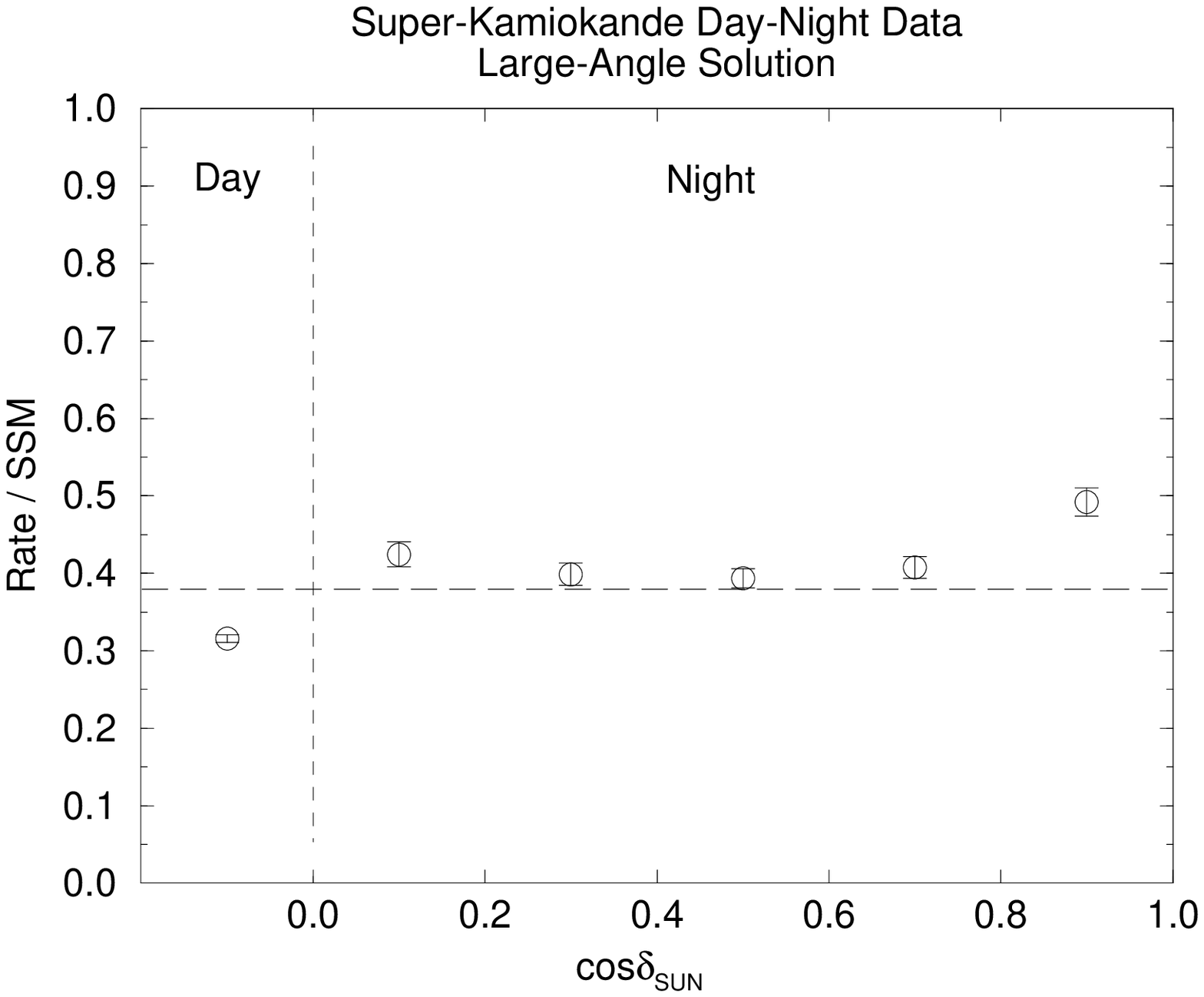}}
\end{tabular}
\end{center}
\caption{
%
%
Same as Fig.~\protect\ref{fig_dn_sno}, but for Super-Kamiokande.  The
error bars are the statistical uncertainties assuming a total of 8,000
events (equivalent to one year of operation).
Taken from Ref.~[\protect\citenum{HL-MSW-analysis}].
}
\label{fig_dn_superk}
\end{figure}

\begin{figure}[pbth]
\postscript{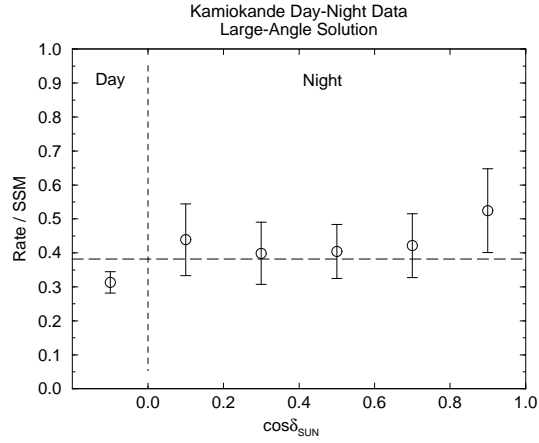}{0.45\hsize}                    
\vspace{4ex}
\caption{
%
%
Same as Fig.~\protect\ref{fig_dn_sno}, but for Kamiokande.  The error
bars are the statistical uncertainties assuming a total of 200 events.
Taken from Ref.~[\protect\citenum{HL-MSW-analysis}].
}
\label{fig_dn_kam}
\end{figure}

The large-angle solution of the combined analysis predicts a
night-time enhancement of the signal which should be observable in SNO
and Super-Kamiokande.  The predictions of the day-night effect in the
small-angle (nonadiabatic) and the large-angle solutions are shown in
Fig.~\ref{fig_dn_sno} and \ref{fig_dn_superk}.  In fact the
enhancement is such a large effect that it might be observable in the
combined data of Kamiokande II and III (Fig.~\ref{fig_dn_kam}).

An interesting possibility of detecting the day-night effect for the
small-angle (nonadiabatic) solution was carefully investigated by
Baltz and Weneser recently \cite{Baltz-Weneser-94}.  It was pointed
out that, for the best fit parameters in the small-angle
(nonadiabatic) region, the $^8$B neutrinos can resonate in the core of
the Earth, which corresponds to the last bin in Figs.~\ref{fig_dn_sno}
(a) and \ref{fig_dn_superk} (a).  Separating those signals, it was
concluded that Super-Kamiokande should be able to see the Earth effect
at about $2\sigma$ in a year.  It is still an open question whether
additional information from the energy spectrum distortions enhances
the signals, and whether the effect is therefore observable with
definitive statistics above the background.

\begin{figure}[bp]
\begin{center}
\begin{tabular}{c}
\setlength{\epsfxsize}{\mswsize}
\subfigure[Super-Kamiokande]{\epsfbox{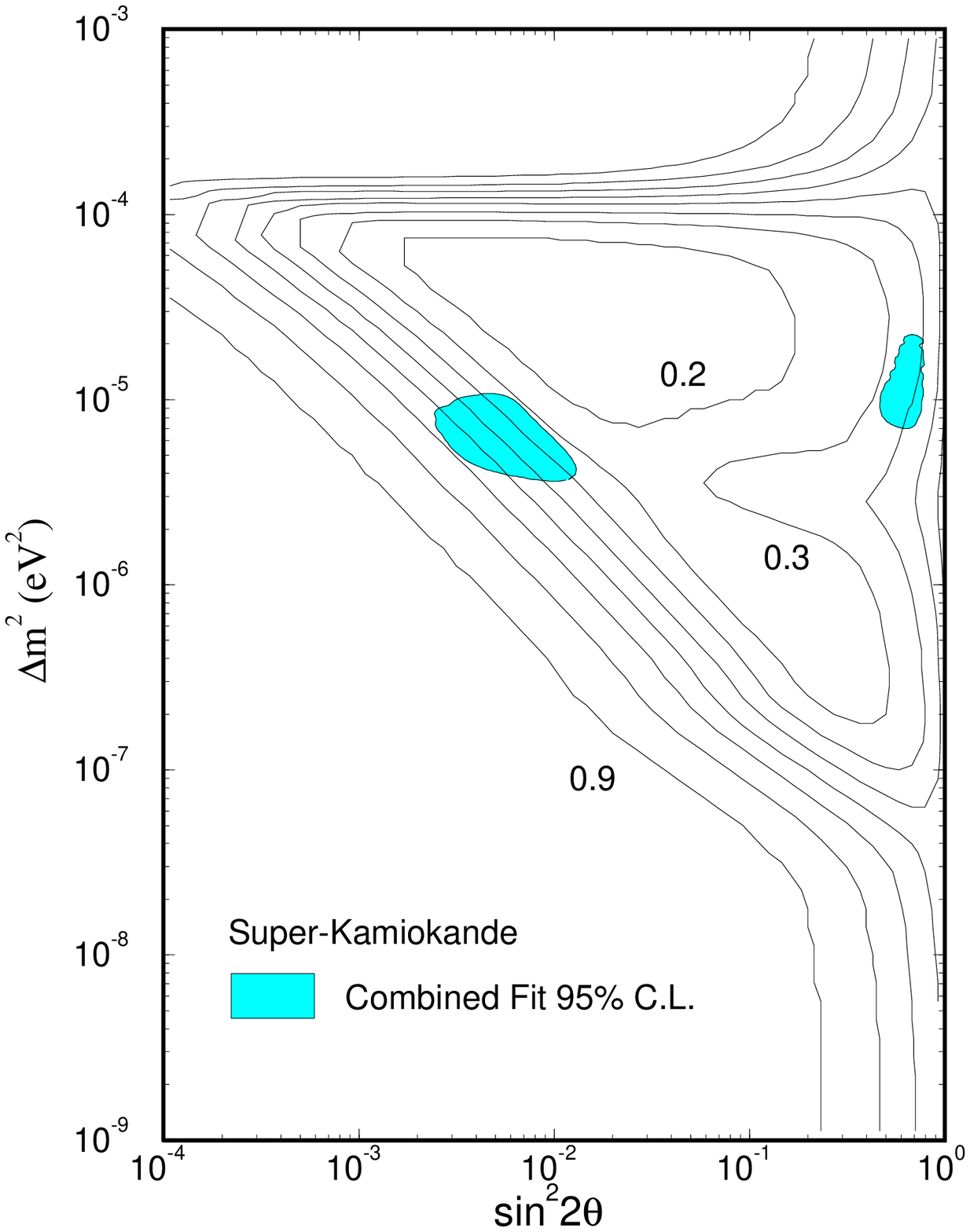}}
\hspace{3em}
\subfigure[BOREXINO]{\epsfbox{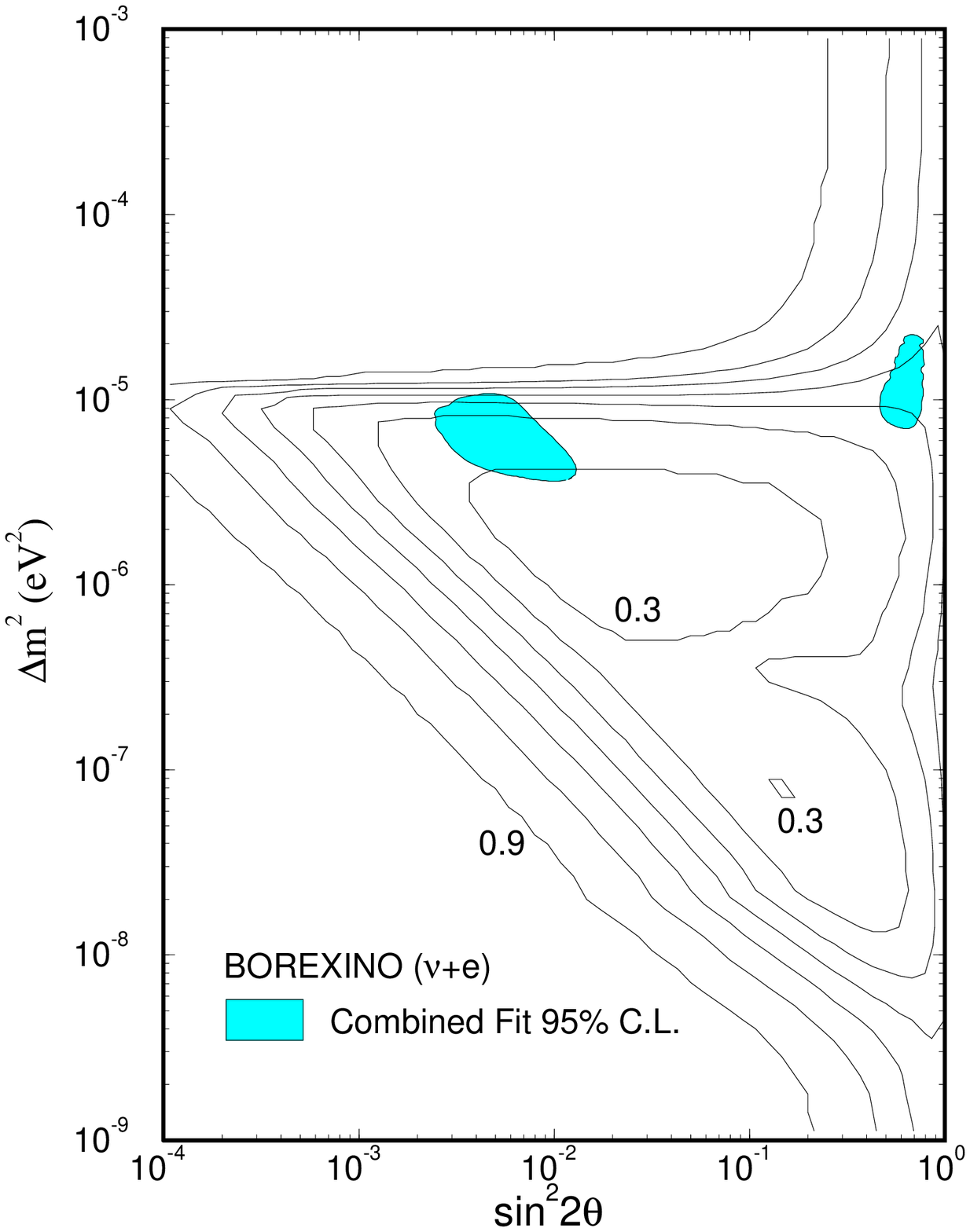}}
\end{tabular}
\end{center}
\vspace{-1ex}
\caption{
%
%
The contours of the signal-to-SSM ratio for (a) Super-Kamiokande and
(b) BOREXINO, along with the global MSW solutions (from present data)
obtained in the Bahcall-Pinsonneault SSM.  This is an updated result
of Ref.~[\protect\citenum{HL-MSW-analysis}].
}
\label{fig_c_future}
\end{figure}

\subsubsection{		Summary of MSW Signals		}

The parameter regions allowed by the combined observation are shown
with the contours of the signal-to-SSM ratio in
Fig.~\ref{fig_c_future} for Super-Kamiokande and BOREXINO, assuming
the Bahcall-Pinsonneault SSM.  Given the high event rates in SNO,
Super-Kamiokande, and BOREXINO, the accurate determination of the
neutrino parameters should be possible.

The various characteristic signals from the global MSW solutions are
summarized in Table~\ref{tab_future-signals-flavor}.  Also listed is
the vacuum oscillation solution ($\Delta m^2 \sim 8 \times 10^{-11}\,
\mbox{eV}\, ^2$ and $\sin^22\theta \sim 0.8$), which is allowed by the
data, but requires a fine-tuning between the Sun-Earth distance and
the neutrino parameters.  It is stressed that the significance of
establishing neutrino oscillations by the CC/NC measurement in SNO
depends on the solar model.  On the other hand, the effects of
spectrum shape distortions, the day-night effect, and seasonal
variations are free from solar model uncertainties and theoretically
clean.
An equivalent list for oscillations to sterile neutrinos are shown in
Table~\ref{tab_future-signals-sterile}.

\vspace*{-2.3ex}

\begin{table}[pht]
\begin{center}
\caption{
Characteristic signals expected in the global MSW solutions in flavor
oscillations.  The second large-angle solution is strongly disfavored
by the data, and the solution only appears at 99\% C.L.  The vacuum
oscillation solution is also listed.  Whether SNO can definitively establish
neutrino oscillations by the CC/NC measurement is solar model
dependent.
}
\vspace{1ex}
\label{tab_future-signals-flavor}

\small
\begin{tabular}{ l c c c c}
\hline
\hline
	& \multicolumn{4}{c}{Oscillations to $\nu_{\mu, \tau}$ } \\
\hline
	           & Nonadiabatic& Large-Angle (I)& Large-Angle (II) & Vacuum\\
\hline
CC/NC	           &   $\surd$	 &   $\surd$     &  $\surd$    &  $\surd$    \\
Spectrum Distortion &  $\surd$   &               &             &  $\surd$    \\
Day-Night Effect ($^8$B) &  ?    &   $\surd$     &             &             \\
Day-Night Effect ($^7$Be)&       &               &  $\surd$    &             \\
Seasonal Variations &            &               &             &  $\surd$    \\
\hline
\hline
\end{tabular}
\normalsize
\end{center}
\end{table}

\nopagebreak
\vspace{-7.5ex}
\begin{table}[pbh]
\begin{center}
\caption{
Characteristic signals expected in the global MSW solutions for
sterile neutrinos.  The large-angle solutions (both (I) and (II)) and
the vacuum solutions are strongly disfavored by the data and excluded
for sterile neutrinos.  The large-angle (I) solution is independently
excluded by the big-bang nucleosynthesis constraint
\protect\cite{sterile,Shi-Schramm-Fields}.
}
\label{tab_future-signals-sterile}
\vspace{1ex}
\small
\begin{tabular}{ l c c c c}
\hline
\hline
	& \multicolumn{4}{c}{Oscillations to $\nu_s$}                        \\
\hline
	           & Nonadiabatic& Large-Angle (I)& Large-Angle (II) & Vacuum\\
\hline
CC/NC	           &  		 &  		 &	       &             \\
Spectrum Distortion&   $\surd$   &               &             &  $\surd$    \\
Day-Night Effect ($^8$B)&  ?     &   $\surd$     &             &             \\
Day-Night Effect ($^7$Be)&       &               &  $\surd$    &             \\
Seasonal Variations&             &               &             &  $\surd$    \\
\hline
\hline

\end{tabular}
\normalsize
\end{center}
\end{table}

\section{		Summary and Wish List		}
\label{sec_summary}

\subsection{		Summary				}

Various theoretical models were confronted with the solar neutrino
observations as possible solutions of the solar neutrino deficit.  The
astrophysical solutions are strongly disfavored by the data: (a) The
SSM is excluded by the each of the experiments.  (b) The model
independent analysis concludes that the combined results from
Homestake and Kamiokande exclude any astrophysical solutions.  (c)
Even if the Homestake data were discarded, we have no realistic solar
model that simultaneously explains the Kamiokande and the gallium
results within the experimental uncertainties.

The MSW mechanism, on the other hand, provides a complete description
of the data, without discarding any of the observations.  The parameter
space allowed by the data is constrained to
\begin{eqnarray}
& \mbox{Nonadiabatic:} &\; \Delta m^2 \sim 10^{-5} \, \mbox{eV}\, ^2
                           \mbox{ and } \sin^22\theta \sim 0.007  \\
& \mbox{Large-angle:}  &\; \Delta m^2 \sim 10^{-5} \, \mbox{eV}\, ^2
	                   \mbox{ and } \sin^22\theta \sim 0.8.
\end{eqnarray}
There is a third allowed region for $\Delta m^2 \sim 10^{-7} \,
\mbox{eV}\, ^2$ and $\sin^22\theta \sim 0.8$, but only at 99\% C.L.
For oscillations to sterile neutrinos, the solution is allowed only in
the nonadiabatic region.

The obtained parameters are sensitive to the solar model
uncertainties.  When nonstandard core temperatures are allowed, the
data constrain $T_C = 1.02 \pm 0.02$ (1$\sigma$), consistent with the
SSM.  A wider MSW parameter space is possible in nonstandard solar
models.

The predictions for the CC/NC ratio, spectrum distortions, and
day-night effect in SNO and Super-Kamiokande were discussed in detail.
It is stressed that the MSW prediction for the CC/NC ratio is solar
model dependent, and a smaller $^8$B flux can make the difference
between the MSW solutions and no-oscillations ambiguous.

\subsection{		Wish List			}

The solar neutrino experiments have provided us an opportunity for a
better understanding of the Sun as well as a possible discovery of new
neutrino physics.  There will be more data soon, and much work is
still needed involving experimentalists, astrophysicists, nuclear
physicists, as well as particle physicists.  The INT workshop allowed
us to convene and discuss this inter-disciplinary topic, and, taking
advantage of the opportunity, I will address my hopes from a particle
physicist point of view:

\vspace{1ex} \noindent
{\bf From experimentalist friends:}  \nopagebreak
\begin{itemize}
\item	{\it Uncertainty correlation matrix be published.}
	In constraining the parameters in the MSW analysis, the energy
	spectrum and the time-divided data are very informative.  From
	published papers, however, it is often not clear which of the
	systematic uncertainties are correlated among data points.  In
	high-energy experiments such as in LEP, it is becoming more
	common to publish the correlation matrix, and I hope this
	trend will be followed in solar neutrino experiments.
\end{itemize}

\noindent
{\bf From astrophysicist friends:}
\begin{itemize}
\item	{\it The $S_{17}$ uncertainty be reevaluated.}
	It is my belief that the current uncertainty obtained in
	Ref.~[\citenum{Johnson-etal}] and used in the
	Bahcall-Pinsonneault model is underestimated.  A larger
	uncertainty does not solve the solar neutrino problem;
	however, it does change the obtained MSW parameter space and
	affects the MSW predictions for the future experiments,
	especially the CC/NC ratio in SNO.

\item	{\it A search for nonstandard solar models consistent with
	the Kamiokande and gallium data be pursued.}  There is no way
	one can reconcile the Homestake and Kamiokande data by
	astrophysical explanations.  Even discarding the Homestake
	results entirely, we have no realistic solar model consistent
	both with the Kamiokande and gallium data, but more studies
	are needed.  Especially, every nonstandard solar model should
	be confronted with the helioseismology data.  If we cannot
	find a solar model consistent with the Kamiokande, gallium,
	and the helioseismology data after discarding the Homestake
	results, our motivation for new neutrino physics will be
	complete.

\item	{\it Can the neutrino flux and the MSW parameters be constrained
	simultaneously?}  Once neutrino oscillations are established
	in the next generation experiments, the most significant task
	is to constrain the neutrino flux and the MSW parameters
	simultaneously from observations.  This can be complicated
	since the MSW mechanism can distort the neutrino spectrum.
	Either simple relations among the fluxes (like the $T_C$ power
	laws) or Monte-Carlo models like the ones by Bahcall and
	Ulrich but with a wider range of solar models are needed.  It
	would be ideal if the helioseismology data were incorporated
	in such an analysis.

\end{itemize}

\centerline{\bf		  Acknowledgments     		}

This paper is based on the work done in collaboration with Sidney
Bludman and Paul Langacker.  I am grateful to P.\ Langacker for his
careful reading of the manuscript and numerous valuable suggestions.
I thank Institute for Nuclear Theory at the University of Washington
for its hospitality and the Department of Energy for partial support
during the time a part of this work was done.  It is pleasure to thank
the participants of the INT workshop for useful discussions;
especially I have benefited from stimulating discussions with Anthony
Baltz on the Earth effect.  I also thank Eugene Beier for useful
discussions.  This work is supported by the Department of Energy
Contract DE-AC02-76-ERO-3071.



\end{document}